# The sunspot number record supports the existence of Planet 9 and the effect of planetary motion on solar activity.

Ian R. Edmonds

Formerly: Department of Physics, Queensland University of Technology, Brisbane, Australia.

**Email address:**

iredmonds@aapt.net.au

**Abstract:** This paper assesses if the Planet 9 hypothesis, the existence of a ninth planet, is consistent with the planetary hypothesis, the synchronisation of sunspot emergence to solar inertial motion (SIM) induced by the planets. We show that SIM would be profoundly affected if Planet 9 exists and that the hypothesised effect of SIM on sunspot emergence would be radically different from the effect of SIM due to the existing eight planets. We compare the spectral and time variation of Sun to barycentre distance, $R_B$, calculated for both the eight and nine planet systems, with variation of sunspot number (SSN). Including Planet 9 improves spectral correlation and time coherence between $R_B$ and SSN in the decadal, centennial and millennial time range. Additionally, as the variation of $R_B$ is sensitive to Planet 9 parameters, longitude and period, it is possible to tune both parameters to SSN variation and obtain new estimates of the Planet 9 parameters independent of astronomical observations. We develop a mechanism for the influence of SIM on SSN that provides an explanation of the consistency between SIM, calculated with Planet nine, and records of reconstructed SSN.

## 1. Introduction

The discovery of Neptune in 1846 was facilitated by observations suggesting the orbit of Uranus was perturbed by an unknown planet. Subsequent calculations, based on the perturbations, provided an estimate of Neptune's location in the sky and shortly after it was observed directly. Recent astronomical observations of the orbits several Kuiper Belt objects have led to the hypothesis that a ninth planet, far more distant than Neptune, exists and is perturbing the orbits, Brown and Batygin (2021). Their calculations, based on the apparent clustering of the longitude of perihelion of the objects, have provided estimates of the mass, semi major axis, eccentricity, and inclination of the hypothetical planet, known as Planet 9, and an indication that the likely location of the planet is in the vicinity of Orion's Shield. Planet 9 has not been directly observed and the hypothesis that it exists is controversial. The absence of direct observation is due to the extreme difficulty of observing so distant a planet with existing telescopes. However, the controversy arises mainly from the fact that the hypothesis relies on the clustering in longitude of perihelion of just six objects out of a much larger population of objects and may be the result of selection bias, Napier et al (2021). Other than this clustering in longitude of perihelion of a few objects the additional supporting evidence appears to be limited to an apparent trend in the declination of Pluto over the past 20 years, Holman and Payne (2016). However, the trend favours a planet either more massive or closer than the planet hypothesised by Brown and Batygin (2021) or possibly another planet, closer to Pluto, in addition to Planet 9; a result that tends to increase rather than decrease the controversy. Clearly, in the continuing absence of direct observation, e.g. Naess et al (2021), other supporting evidence would be welcome.

A possible source of support for the Planet 9 hypothesis could come from the planetary hypothesis, Charbonneau (2010), that there is a causal link between the cycles of sunspot number (SSN), such as the ~11 year Schwabe, ~ 88 year Gleissberg, ~ 1000 year Eddy and the ~ 2400 year Hallstatt cycles and planetary motion. The planetary hypothesis has been developed, with limited acceptance, e.g. (Jose 1965, Fairbridge and Shirley 1987, Charvatova 2000, Abreu et al 2012, Scafetta 2012, Scafetta and Willson 2013, Wilson 2013, Scafetta et al 2016, Charvatova and Hejda 2014, Cionco and Soon 2015, Stefani et al 2020B). The more conventional view of the origin of cycles in SSN is that sunspot emergence is irregular and is associated with the effect, on the solar dynamo, of random (stochastic) hydro-magnetic flows in the convective zone of the Sun, (Usoskin et al 2009, Charbonneau 2020).  The origin of the cycles or quasi-cycles in sunspot emergence is controversial partly due to the extended range of sunspot cycle periodicity. Examples of longer term SSN cycle variability observed include ~60, ~88, ~104, ~150, ~208, and ~506 year periods, in the centennial periodicity range, (Abreu et al 2012, Tan and Cheng 2012), and the ~1000 year Eddy and ~2400 year Hallstaat cycles in the millennial periodicity range, (Eddy 1976, Steinhilber et al 2012, Usoskin et al 2016, Scafetta et al 2016). Various theories have been proposed to account for this range of periodicity in sunspot emergence: some theories based on purely interior mechanisms e.g. (Dikpaki and Gilman 2006, Zaqarashvili et al 2010, Tobias et al 2006, Weiss and Tobias 2016, Beer et al 2018, Liang et al 2019, Karak and Choudhuri 2013) and some theories based on planetary synchronised mechanisms e.g. (Abreu et al 2012, Wolff and Patrone 2010, Stefani et al 2020B). All previous planetary synchronism mechanisms have been based on the eight planet solar system. No study has investigated the possible effect of a distant ninth planet on planetary synchronised sunspot emergence.

The planetary synchronism of sunspot emergence due to the cyclic motion of Sun about the solar system barycentre, e.g. (Jose 1965, Landscheidt 1999, Wolff and Patrone 2010, McCracken et al 2014) would be profoundly affected, in terms of periodicity and time variation, by a ninth planet. The simple reason for this is that, as displacement of the Sun from the barycentre is proportional to the product of planet mass and planet distance from the barycentre, a very distant planet can have a large effect. Jose (1965) calculated, for the years 1653 to 2060, the distance, $R_B$, between the Sun and the solar system barycentre for the eight planet system and was able to show a reasonable correlation between $R_B$ and the record of SSN available at that time, up to 1964, provided the SSN was signed, i.e. that each second cycle of sunspot emergence was given a negative sign. However, comparison of the timing of SIM with the occurrence of more recent decadal solar cycles proved unconvincing, Fairbridge and Shirley (1987). As a result subsequent studies of the connection between SIM and SSN shifted to comparisons of SIM with SSN on the centennial scale, specifically the occurrence of grand solar minima and maxima, e.g. (Fairbridge and Shirley 1987, Charvatova 2009, McCracken et al 2014, Cionco and Soon 2015). With eight planets, relating SIM to SSN on the centennial and millennial scale has proven problematic due to $R_B$ being essentially constant when averaged over centennial and millennial time scales, (Fairbridge and Shirley 1987, Cionco and Pavlov 2018), a fact confirmed in Figure 1 of this paper. Nevertheless, workers have been able to find the ~ 2400 year cycle in the patterns of SIM, (Fairbridge and Shirley 1987, Shirley 2009), in differences in the ordered and disordered states of SIM, (Charvatova 2009, McCracken et al 2014, Charvatova and Hedja 2014), and in very small variations in the average ellipticity of the solar orbit about the barycentre, Scafetta et al (2016).

In this paper we consider the possibility that relating SSN to SIM has proven difficult because the calculation of SIM has not included the effect of Planet 9. If Planet 9 exists, the projected mass, $m_9$, is about seven Earth masses, and the projected semi-major axis is about 380 AU, Brown and Batygin (2021). The displacement of the Sun from the barycentre is proportional to the product $m_P r_P$ where $m_P$ is the planet mass and $r_P$ is the distance of the planet from the Sun. For Jupiter, $m_J$ = 318$m_E$ and $r_J$ ~ 5.2 AU and for Planet 9, $m_9$ ~ 7$m_E$, and $r_9$ ~ 380 AU. The ratio of the two displacements is 7x380/318x5.2 = 1.6. So even though Planet 9 is very distant and orbiting very slowly, the effect on SIM about the barycentre will be larger than the effect of Jupiter. Therefore, it is perhaps timely to consider if the inclusion of Planet 9 in calculations of SIM improves the frequency and time relationship of SIM to solar activity. The corollary, that if a nine planet formulation of SIM proves more consistent with the frequency and time dependence of solar activity than the current eight planet formulation, that in itself would provide additional indirect evidence for the existence of Planet 9.

Section 2 of the paper briefly discusses the current knowledge of Planet 9 and outlines the data sources and the simplified method of calculating the time variation of the SIM used in the paper. Section 3 demonstrates that SIM with Planet 9 included provides a better fit to the decadal variation in solar activity than SIM without Planet 9. Section 4 uses the observed sunspot activity from 1610 to the present to show that recent grand solar minima are associated with large decreases in centennial scale averages of $R_B$. Section 5 demonstrates that, in the millennium scale variation of SIM with Planet 9 included, several significant components, the Hallstatt, Gleissberg, 60 and 30 year cycles, emerge and demonstrates that these components are absent in SIM without Planet 9. Additionally, it is shown that low frequency spectral components of SIM are sensitive to the orbital period of Planet 9 and, consequently, the orbital period can be tuned to fit the low frequency component of SIM to the Hallstatt cycle in solar activity. Section 6 demonstrates the phase modulation of the Jose cycle in the reconstructed SSN. Section 7 introduces a mechanism for the influence of SIM on meridional flow in the convective region of the Sun and, ultimately, on SSN. Section 7 is a conclusion.

## 2. Methods and data sources

**2.1 Prior evidence of Planet 9.** Evidence for the existence of Planet 9 has been accumulating for about 20 years, (Brown et al 2004, Trujillo and Sheppard 2014, Batygin 2016, Bailey et al 2016, Batygin et al 2019). Analysis of the evidence, the anomalous orbits of some Kuiper belt objects, suggests the existence of a new planet of mass ~ 7 Earth masses, in an orbit of eccentricity ~ 0.3 with a semi major axis of ~ 380 AU inclined at ~ 15 degrees to the ecliptic plane, Brown and Batygin (2021). However, the parameter estimates, based just on the anomalous clustering of distant objects, remain very uncertain, Brown and Batygin (2021).

**2.2 Data sources.** Planet heliographic longitudes on January 01, 1965, were obtained from https://omniweb.gsfc.nasa.gov/coho/helios/heli.html. Group sunspot numbers 1610 to 2015, Svalgaard and Schatten (2016), were obtained from https://svalgaard.leif.org/research/gn-data.htm. Reconstructed SSN -6755 to 1885 was obtained from https://www2.mps.mpg.de/projects/sun-climate/data/SN_composite.txt

**2.3 Method of calculating SIM with Planet 9 included.** In view of the limited accuracy of the orbital parameters of Planet 9, in this article, SIM is calculated using a model where all the planets move about the Sun in circular orbits in the ecliptic plane. This is a good approximation for the known planets that have a significant effect on SIM, i.e. Jupiter, Saturn, Uranus and Neptune, as the orbits these planets are nearly circular and have low inclination to the ecliptic plane. As shown in Figure 1 below, the SIM, calculated using circular orbits for the eight known planets, is scarcely distinguishable from SIM calculated using exact planet orbits based ephemeris data. However using a circular orbit for Planet 9 is an approximation as its orbit is projected to be eccentric, $\varepsilon$ ~ 0.3, and inclined to the ecliptic, i ~ 15$^o$, Brown and Batygin (2021). Including eccentricity and inclination of Planet 9 in the calculation of SIM would greatly increase the complexity of the calculation and is outside the scope of this early investigation. Spectral analysis will be an important investigative tool in this paper and for basic spectral analysis like FFT equal time intervals are essential. With circular orbits, as used in this paper, equal time intervals occur naturally in the calculation of SIM whereas for eccentric orbits equal angular intervals occur naturally and the conversion to equal time intervals is complex. The orbital parameters of the planets used in the calculation are given in Table 1.

**Table 1. Parameters of the planets** Planet 9 table 1 planet parameters.docx

| Planet | Mass in mass Earth units, $m_i$ | Radius, AU, $r_i$ | Period, days $T_i$ (years) | HGILong 01/01/1965, $L_i$ |
|---|---|---|---|---|
| Jupiter | 318 | 5.2 | 4332 (11.8) | 340.2 |
| Saturn | 95 | 9.5 | 10759 (29.4) | 260.6 |
| Venus | 0.815 | 0.72 | 224.7 (0.615) | 146.2 |
| Earth | 1 | 1 | 365.256 (1) | 24.9 |
| Neptune | 17.15 | 30.07 | 60195 (164.8) | 152.6 |
| Uranus | 14.53 | 19.19 | 30688 (84.0) | 86.7 |
| Mercury | 0.055 | 0.387 | 87.969 (0.241) | 85.2 |
| Mars | 0.107 | 1.524 | 687 (1.88) | 63.8 |
| Planet 9 | 7 | 380 | 2,700,000 (7400) | 60 (fitted) |
| Sun | 333,000 | | | |

The orbital period of Planet 9, $T_9$ = 2,700,000 days, 7,400 years, is obtained from the value of the semi-major axis, $a_9$ = 380 AU, the best estimate by Brown and Batygin (2021), and the use of Kepler's 3rd Law, $T^2/a^3$ = constant. If T is measured in years and a is measured in AU, the constant = 1.

The time variation of the coordinates, $(x_i,y_i)$, of the ith planet relative to the Sun as origin at (0,0) are calculated using

$x_i = r_i\cos(\omega_i t + \phi_i), \quad y_i = r_i\sin(\omega_i t + \phi_i),$ (1)

where $r_i$ is the orbital radius of the ith planet, the angular frequency $\omega_i = 2\pi/T_i$, the phase angle in radians, $\phi_i = (\pi/180)L_i$, and $L_i$ is the heliographic inertial longitude of the planet in degrees on January 01, 1965.

The coordinates, $(x_{PCM}, y_{PCM})$, of the planetary centre of mass (PCM) relative to the Sun are given by

$x_{PCM} = \Sigma m_i x_i / \Sigma m_i, \quad y_{PCM} = \Sigma m_i y_i / \Sigma m_i$ (2)

and the distance between the Sun and the planetary centre of mass, $r_{PCM}$, is given by

$r_{PCM} = (x_{PCM}^2 + y_{PCM}^2)^{1/2}$ (3)

The distance between the Sun and the barycentre, $R_B$, is

$$R_B = r_{PCM}[\Sigma m_i/\{\Sigma m_i + m_{SUN}\}] \qquad (4)$$

The Sun to barycentre distance is usually expressed as the ratio $R_B/R_{SUN}$ where $R_{SUN}$ is the radius of the Sun, 0.0046 AU.

### 3. Comparing the decadal scale variation of the SSN record and $R_B/R_{SUN}$.

**3.1 The decadal scale variation of $R_B/R_{SUN}$ for the eight and nine planet systems.** Figure 1A shows the time variation of $R_B/R_{SUN}$ for the eight planet system, $m_9 = 0$, and Figure 1B shows $R_B/R_{SUN}$ for the nine planet system when $m_9 = 7m_E$. The most striking feature of Figure 1 is that the frequency of the cycles in $R_B/R_{SUN}$ has doubled when $m_9 = 7m_E$. A second feature is that there is a significant centennial scale variation, ~170 year period, when $m_9 = 7m_E$ that is not apparent when $m_9 = 0$.

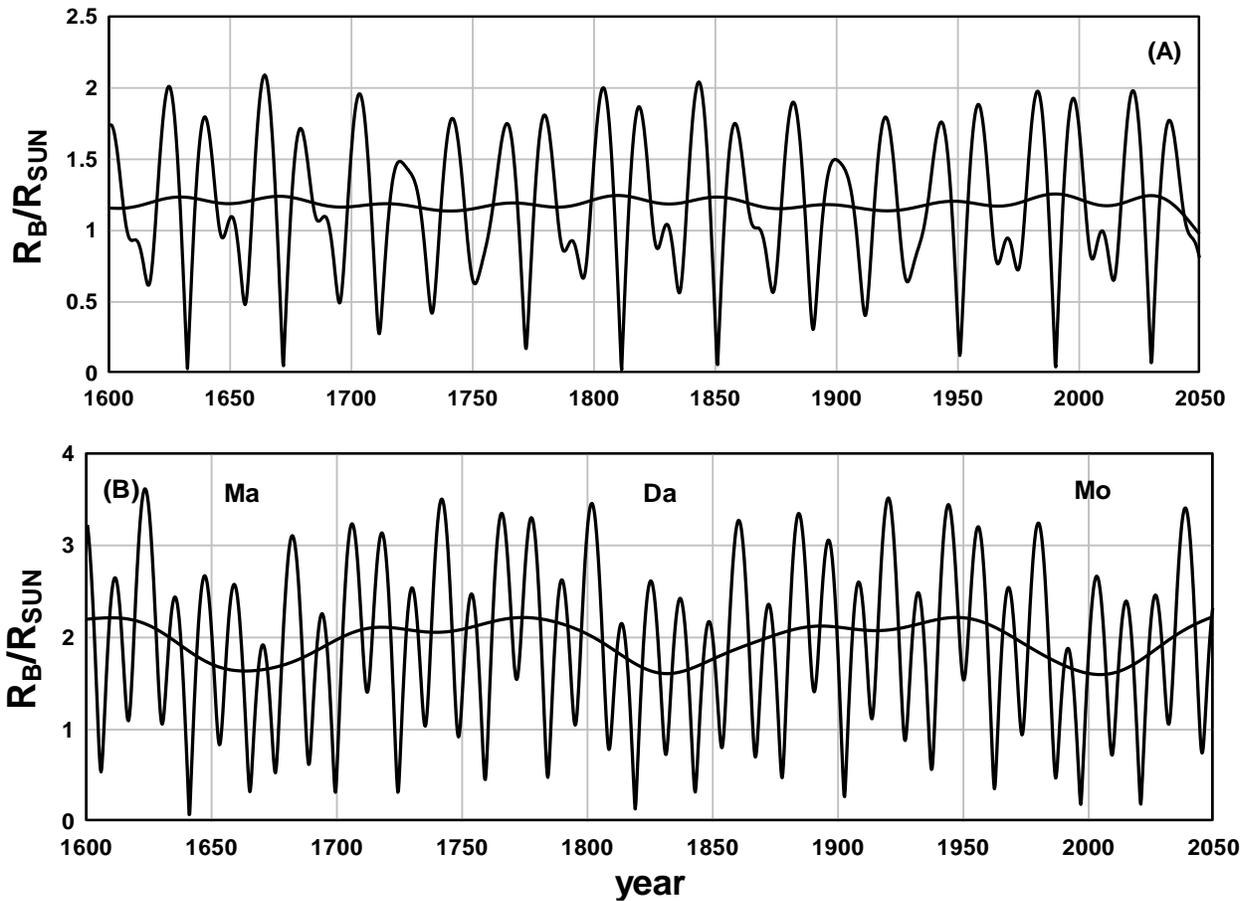

**Figure 1. (A) $R_B/R_{SUN}$ variation for $m_9 = 0$. (B) $R_B/R_{SUN}$ variation for $m_9 = 7m_E$. The graphs illustrate the transition from predominantly bi-decadal periodicity when $m_9 = 0$, to a predominantly decadal periodicity when $m_9 = 7m_E$. Also shown are running averages over 35 years that indicate that large centennial time scale minima occur in $R_B/R_{SUN}$ when $m_9 = 7m_E$ but that no significant centennial scale minima occur in $R_B/R_{SUN}$ for the $m_9 = 0$, the eight planet system. Also labelled, the approximate times of occurrence of the Maunder, Dalton, and Modern grand minima in sunspot number.**

Figure 2 compares the spectral content of the two variations of $R_B/R_{SUN}$ in Figure 1. The spectrum of $R_B/R_{SUN}$ for the eight planet system, $m_9 = 0$, is dominated by a component at 19.5 years, with weaker components at 12.6 and 13.7 years and minor components at 35 and 46 years. Thus the periodicity of the eight planet variation is primarily bi-decadal. The periodicity of $R_B/R_{SUN}$ for the nine planet system is dominated by the component at 11.9 years with weaker components at 8.4, 13.7, 29.6, ~ 86 and ~ 172 years. Thus the periodicity of $R_B/R_{SUN}$ for the nine planet system is primarily decadal. The component at 11.9 years is the synodic period of Jupiter with Planet 9. Similarly the components at 29.6 years, ~86 years and ~172 years are due to Saturn, Uranus and Neptune respectively. The period of a component, given by $T = 1/(1/T_P - 1/T_9)$, is not exactly the orbital period of the planet, $T_P$. This is the result of introducing a very slow moving planet into the solar system.

A feature of $R_B/R_{SUN}$ for the nine planet system is the presence of moderately strong components at ~ 86 years and ~ 172 years. The presence of these longer period components is also clearly evident in the time variation of Figure 1B where minima in the long term average value of $R_B/R_{SUN}$ occur at ~ 170 year intervals with the minima separated by broad maxima. The pattern of broader maxima alternating with sharper minima is clearly the result of the interference of the ~86 year and ~ 172 year cycles. That is, the deep centennial scale minima in Figure 1B occur when the two cycles are both in the negative part of their cycles. The labels Maunder, Dalton and Modern in Figure 1B correspond approximately to the central times of grand solar minima and the times clearly align with the centennial scale minima in $R_B/R_{SUN}$.

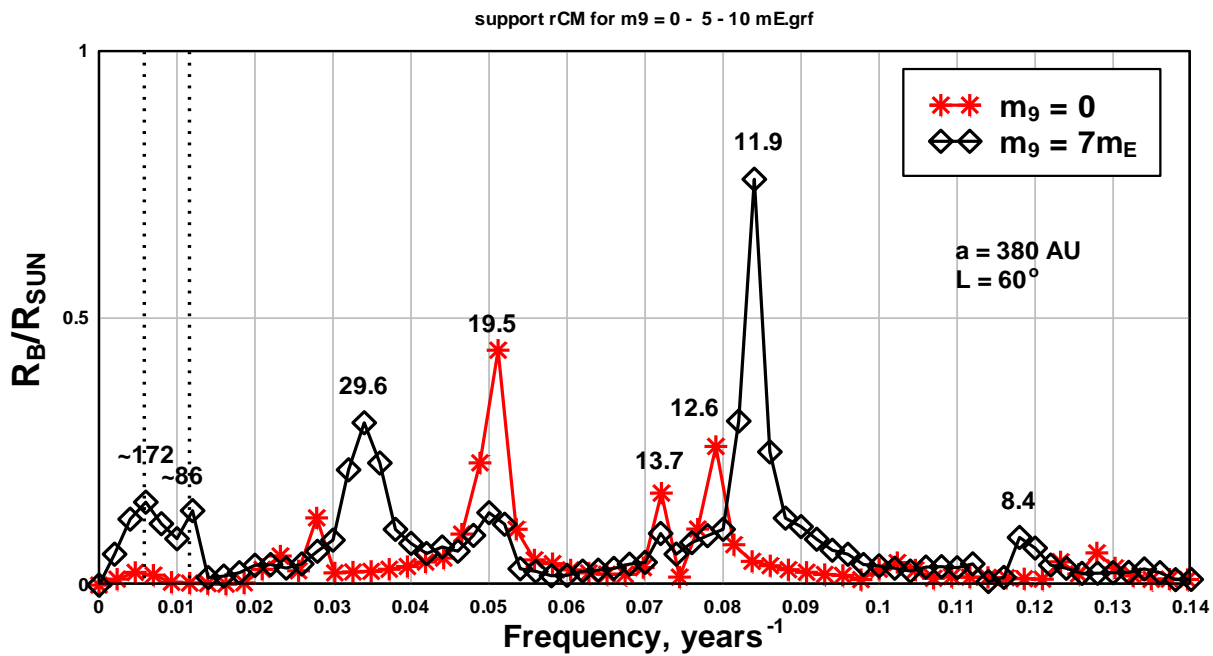

Figure 2. The spectral content of $R_B/R_{SUN}$ for the eight planet solar system, (red stars), and $R_B/R_{SUN}$ for the nine planet system, (black diamonds). For the eight planet system progressively stronger peaks occur at 13.7, 12.6,

and 19.5 years, with weak peaks at 36 and 45 years. When $m_9 = 7m_E$ significant peaks occur at ~172, ~86, 29.6, 11.9 and 8.4 years, with the dominant peak at 11.9 years.

The transition from the primarily bi-decadal SIM for the eight planet system to a primarily decadal SIM for the nine planet system is quite dramatic so it is interesting to follow how this comes about. Figure 3 compares the orbits of the Sun about the barycentre for the eight planet and the nine planet systems during the interval 1890 to 1913, centred on the year 1900, see Figures 1A and 1B.

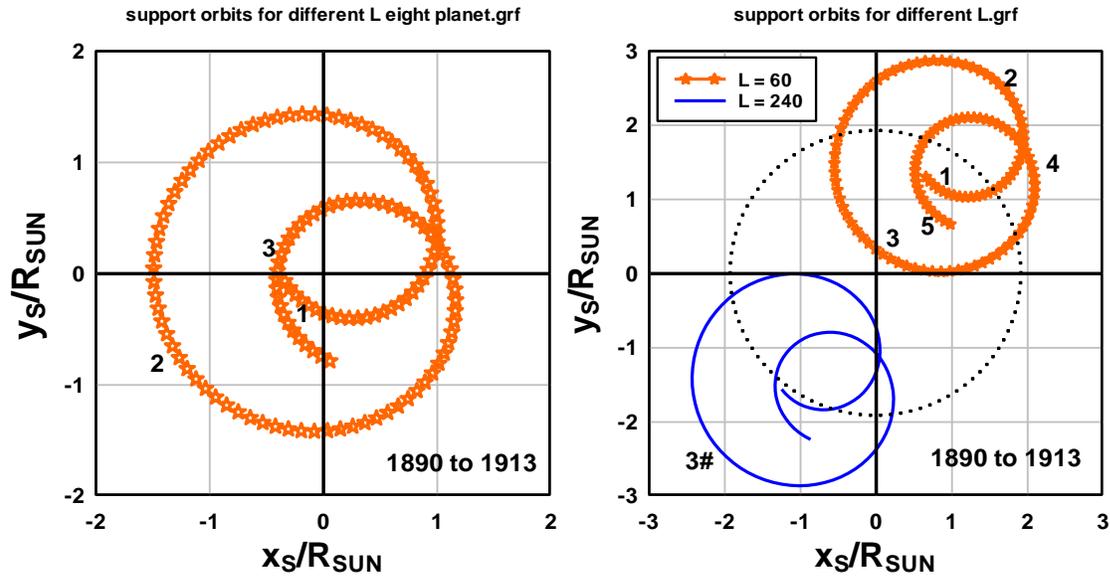

Figure 3. (LHS diagram): Orbit of the Sun about the barycentre between 1890 and 1913 for the eight planet system. This time interval is centred on the year 1900 in Figure 1A. Three extremes in Sun to barycentre distance occur: 1, minimum; 2, maximum; and 3, minimum; corresponding to a ~ 20 year periodicity. (RHS diagram): Orbits of the Sun for the nine planet system in the same time interval centred on year 1900 in Figure 1B. Orbits for two Planet 9 longitudes, L = 60° and L = 240° measured anticlockwise relative to the positive X axis are shown. For the case where L = 60° five extremes occur: 1, minimum; 2, maximum; 3, minimum; 4, maximum; and 5, minimum; corresponding to a ~ 11 year periodicity in Sun to barycentre distance. The dotted circle represents the approximate long term average value of $R_B/R_{SUN}$ over the time interval of one circular orbit of Planet 9, ~ 7000 years. Note that, in Section 5, we show that the long term average value of $R_B/R_{SUN}$ actually varies by approximately 2.5%, from a maximum of 1.95 to a minimum of 1.90 with a period of ~2400 years, the Hallstatt cycle period.

In the left hand diagram of Figure 3 the first minimum in $R_B/R_{SUN}$ occurs at 1 followed by a maximum at 2 and another minimum at 3, with about 20 years between minima, i.e. ~bi-decadal periodicity. The right hand diagram shows the Sun orbit about the barycentre when $m_9 = 7m_E$ for two values of Planet 9 longitude, $L_9 = 60°$ and $L_9 = 240°$. The heliographic inertial longitude, L, is measured anticlockwise from the positive x axis. For the $L_9 = 60°$ case there is a first minimum at 1, a first maximum at 2, a second minimum at 3, a second maximum at 4, and a third minimum at 5, with about 10 years between minima, i.e. ~ decadal periodicity. The phase of the ~ decadal cycle is sensitive to Planet 9 longitude. This is illustrated for the case of $L_9 = 240°$ where the orbital pattern due primarily to Jupiter, Saturn, Uranus, and Neptune has moved from the upper right hand quadrant to the lower left hand quadrant of the diagram. The minimum in $R_B/R_{SUN}$ at 3 occurring at about 1901 is now a maximum in $R_B/R_{SUN}$ at 3# occurring at about 1901. It is apparent that when the longitude of Planet 9 changes by 180° the phase of the short, ~ decadal, cycle in $R_B/R_{SUN}$ also changes by 180°.

**3.2 Comparing the variations of $R_B/R_{SUN}$ and sunspot number.** The group sunspot number (GSS), Svalgaard and Schatten (2016), extends from 1610 to 2015, Figure 4. By removing the 20 year running average we obtain the primarily decadal variation of GSS. The GSS, before 1700, during the Maunder Minimum is of low accuracy, Svalgaard and Schatten (2016), so here we use GSS in the interval between 1700 and 2015 to compare with the variation of $R_B/R_{SUN}$, Figure 5.

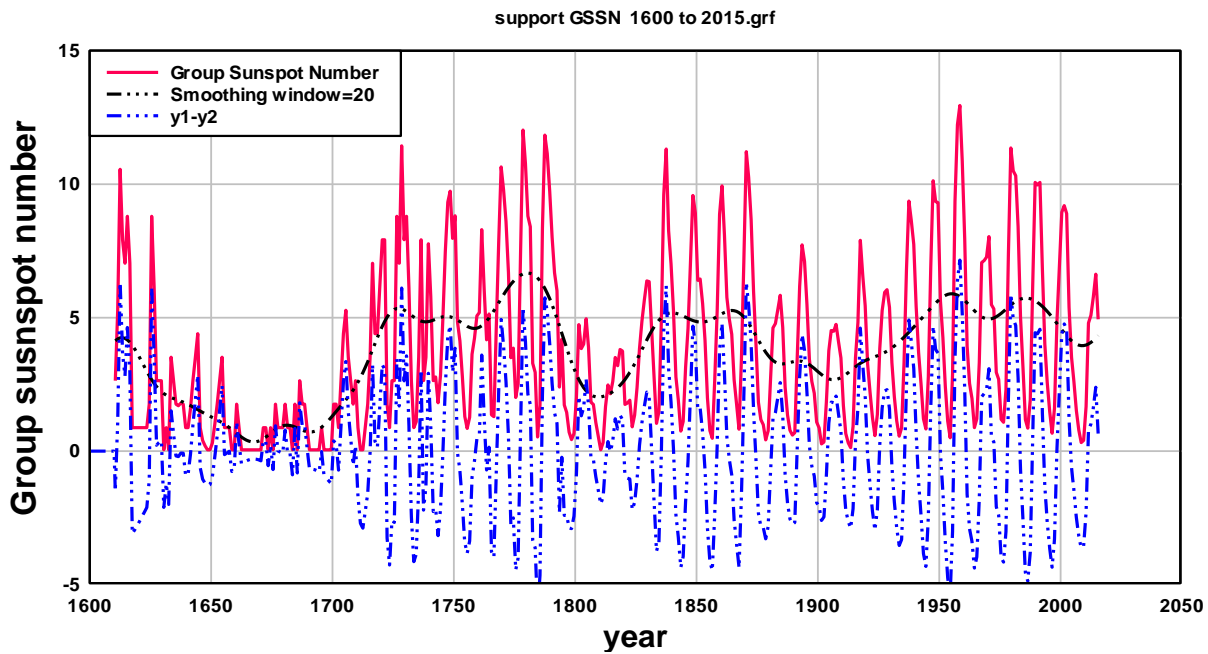

**Figure 4. The group sunspot number (GSS), Svalgaard and Schatten (2016), (red curve) is reduced to the primarily decadal variation, (blue broken curve), by removing the long term variation. The estimated accuracy of the GSS prior to 1700 is low and is not used in the comparison of $R_B/R_{SUN}$ with GSS in Figure 5**

In Figure 5 we use the normalised variations, i.e. the variations divided by their standard deviation, to compare the GSS and $R_B/R_{SUN}$.

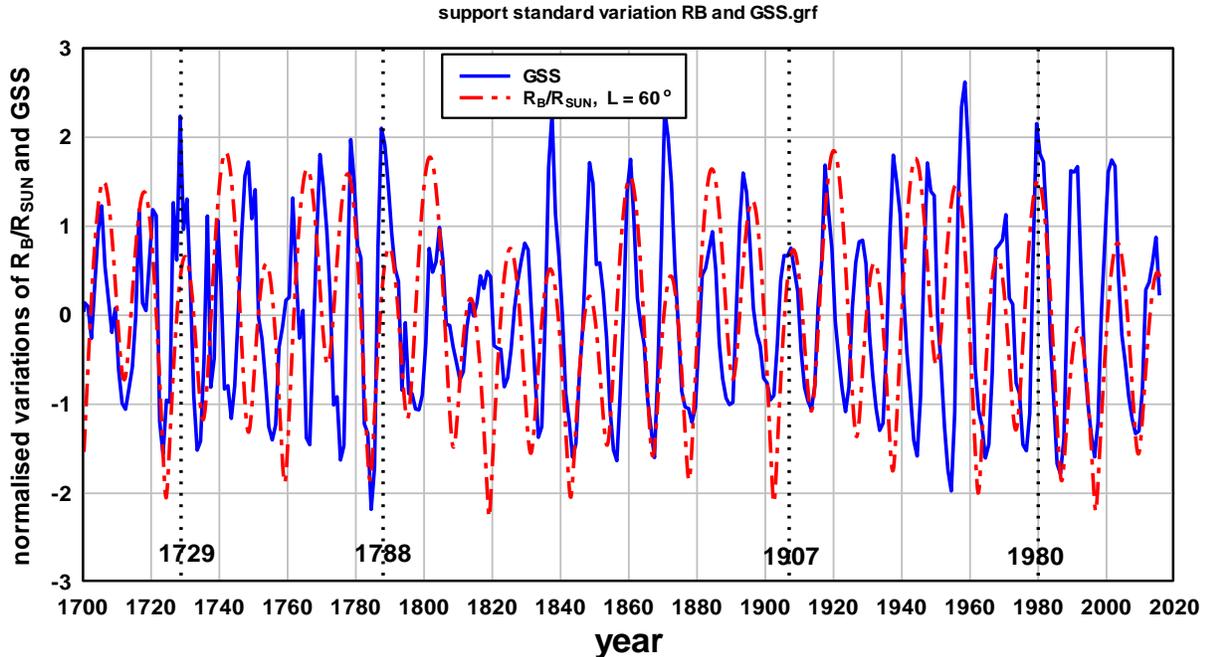

Figure 5. Compares the normalised variations of GSS and $R_B/R_{SUN}$ for the nine planet system. The dotted reference lines mark the intervals when the Schwabe cycle in the GSS jumps in phase by $360°$, i.e. adds an additional cycle, relative to the $R_B/R_{SUN}$ cycle.

There are 28 GSS cycles between 1705.5 and 2014.5. Thus, the average GSS period between the maximum at 1705.5 and the maximum at 2014.5 is 309/28 = 11.03 years. If the hypothesised forcing due to SIM occurs at the dominant periodicity of $R_B/R_{SUN}$ in the decadal time range, i.e. 11.9 years, see Figure 2, we would expect 309/11.9 = 25.96 ~ 26 forcing cycles during this interval. That is, there are 28 GSS cycles compared with 26 SIM cycles during this 309 year long interval. As a result there are two $360°$ phase jumps apparent in Figure 5. The first occurs between 1729 and 1788, (at ~ 1758), and the second occurs between 1907 and 1980, (at ~ 1943). Stephani et al (2020A) observed phase jumps in the Schwabe cycle at around the same times as observed here.

As is evident from Figure 3, the phase of the dominant ~ decadal periodicity in $R_B/R_{SUN}$ is strongly dependent on the longitude of Planet 9. For example the minimum labelled (3) in Figure 3 of the orbital pattern when L = $60°$ becomes a maximum, labelled 3#, in the orbital pattern when L = $240°$. If the correlation coefficient between GSS variation and the $R_B/R_{SUN}$ variation is calculated as the Planet 9 longitude is varied the coefficient passes through a sharp maximum of 0.3 when L = $60°$ as shown in Figure 6. The moderately high correlation coefficient is a result of the fairly close correspondence of the dominant periodicity in $R_B/R_{SUN}$, 11.9 years, and the average periodicity of the GSS between 1700 and 2015, 11.03 years. When the correlation coefficient between GSS and $R_B/R_{SUN}$ for the eight planet system is found by similar analysis for the same time interval the correlation coefficient is 0.034 indicated by the dotted reference line in Figure 6. Note that, in the eight planet calculation, the planet longitudes are not adjustable. A heliographic inertial longitude of L = $60°$ locates Planet 9, at the present time, in the vicinity of the constellation Leo, right ascension ~ 11 hours, declination ~ $+15°$.

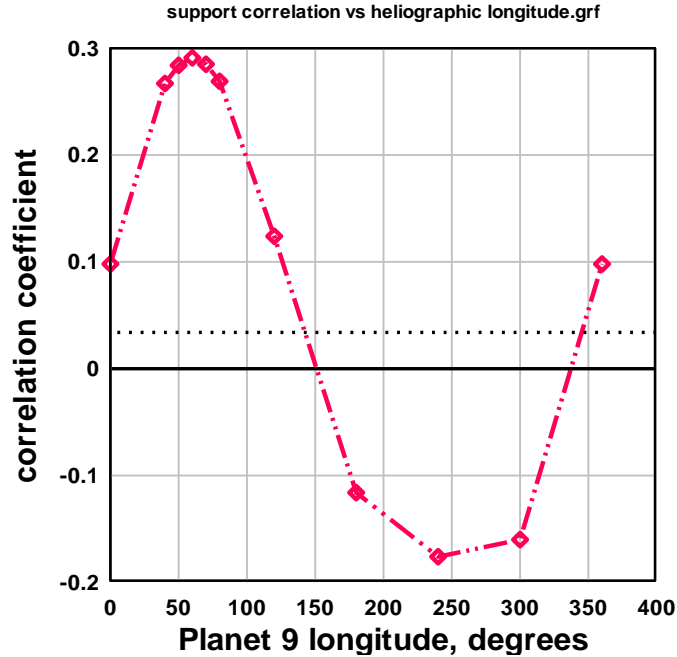

Figure 6. The correlation coefficient between the GSS variation between 1700 and 2015 and the $R_B/R_{SUN}$ variation as the heliographic inertial longitude of Planet 9 varies. The dotted reference line indicates the level of correlation, 0.034, between the GSS and the $R_B/R_{SUN}$ variation for the eight planet system.

4. Comparing the centennial scale variation of SSN and $R_B/R_{SUN}$

Grand solar minima tend to occur in clusters associated with minima in the ~ 2400 year Hallstaat cycle. The most recent minimum of the Hallstatt cycle and most recent cluster of grand solar minima occurred during the last millennium, Usoskin et al (2016). Figure 7, plots the centennial time scale average value of $R_B/R_{SUN}$ for the nine planet system for different values of Planet 9 longitude. Figure 7 shows that the minima evident in the centennial scale variation of $R_B/R_{SUN}$ are sensitive to variation of Planet 9 longitude and that when the Planet 9 longitude is close to 60° the occurrence times of the minima are consistent with the times of occurrence of the most recent grand solar minima, indicated by the vertical reference lines. The times of the recent grand solar minima are taken from the estimates of the central times of grand solar minima, (Usoskin et al 2007, Usoskin et al 2021): Maunder, 1680, Sporer, 1473, Wolf, 1316, and Dalton 1815. The time for the Modern grand minimum, 2030, is an estimate, (Duhau and de Jager 2010, Feynman and Ruzmaikin 2014, Shepherd et al 2014, Morner 2015, Cionco and Soon 2015, Zharkova 2020, Rahmanifard et al. 2022).

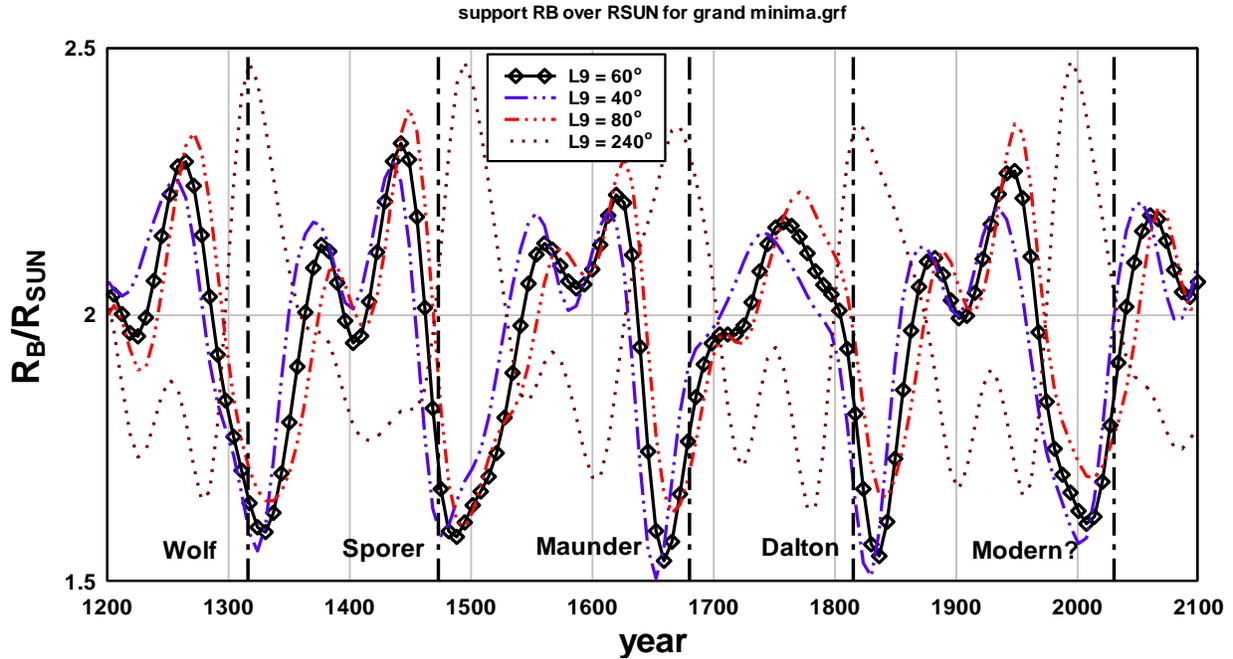

Figure 7. Shows the centennial scale variation of $R_B/R_{SUN}$ between year 1200 and year 2100 for different values of heliographic longitude, HGIL9, of Planet 9. The data, obtained at 6.5 year intervals, has been averaged over 33 years to emphasise the ~ 172 and ~ 86 year components, see Figure 2. The vertical reference lines correspond to dates at the centre of grand solar minima taken from Usoskin et al (2021). The date for the Modern minimum is an estimate.

It is interesting to compare the result in Figure 7 with the analysis of Feynman and Ruzmaikin (2014) who analysed the SSN record from 1700 to 2010 and associated grand solar minima as being due primarily to the ~88 year Gleissberg cycle. Figure 7 clearly indicates that the centennial variation of $R_B/R_{SUN}$ is due to the interference of the ~ 172 year and ~86 year components. As a result $R_B/R_{SUN}$ sharply decreases when the negative phases of the two components interfere constructively. The decreases are separated by broad maxima, occasionally containing a weak minimum, where the positive phase of the ~172 year cycle interferes with negative phase of the ~ 86 year cycle. If the hypothesis of this article that sunspot activity is influenced by variation in $R_B/R_{SUN}$ is correct the result in Figure 7 is also consistent with the recent reconstruction of SSN over the last millennium by Usoskin et al (2021) particularly in respect to the occurrence of grand minima and the occurrence of weak minima between the grand minima. For example, weak solar cycles would be expected at years 1400 and 1900 consistent with the observations of SSN by Usoskin et al (2021).

As noted earlier when Planet 9 is included in the solar system the ~ 172 year component is increased by a factor ~ 10 and the ~ 86 year component, previously not present, appears strongly in SIM, c.f. Figure 2. It is therefore interesting to follow, in terms of SIM about the barycentre, why the introduction of Planet 9 leads to such large centennial scale variations in $R_B/R_{SUN}$, Figure 8.

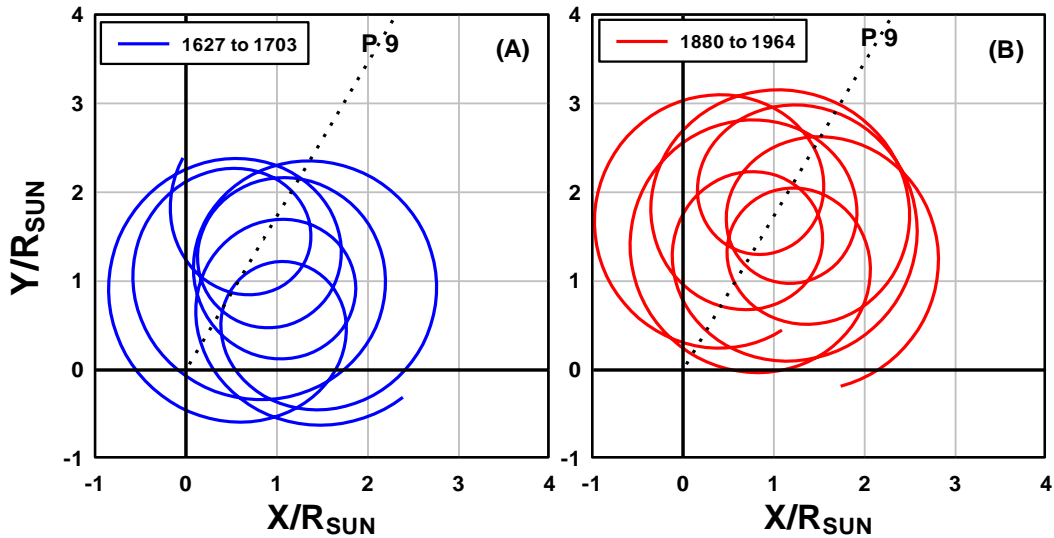

**Figure 8. (A). During a grand solar minimum, here the Maunder Minimum, 1627 to 1703, the known outer planets, Uranus, and Neptune, both lie at longitudes in the lower left quadrant, i.e. are in inferior conjunction, while Planet 9 lies along the longitude, 60°, indicated by the dotted line. As a consequence the orbital pattern of the Sun about the barycentre, indicated in (A), is shifted toward the barycentre and the centennial scale average Sun to barycentre distance is decreased as compared with times when, as in the time interval 1880 to 1964, (B), Uranus and Neptune are near superior conjunction, one in the lower left quadrant and the other in the opposite quadrant, and Planet 9 displaces the orbital pattern away from the barycentre.**

Figure 8 illustrates how the orbital pattern of the Sun about the barycentre, (0,0), varies between a centennial scale grand minimum in $R_B/R_{SUN}$, here 1627 to 1703, and an interval between centennial scale grand minima in $R_B/R_{SUN}$, here 1880 to 1964.

## 5. Comparing the millennial scale variation of reconstructed SSN and $R_B/R_{SUN}$

**5.1 The millennial scale variation of $R_B/R_{SUN}$.** The variability of solar activity over timescales longer than several centuries can be studied by proxy records of solar activity derived from measurements of the concentration of the cosmogenic isotopes, radiocarbon $^{14}C$ in tree trunks and $^{10}Be$ in polar ice, that accumulated due to the effect of cosmic rays, (Vasiliev and Dergachev 2002, Inceoglu et al 2015, Usoskin et al 2016, Vecchio et al 2017, Viaggi 2021). Due to the inverse relationship between cosmic rays and solar activity it was possible to reconstruct the variation of SSN over the approximately 9,000 year Holocene interval, -7000 BC to 2000 AD, (Usoskin et al 2016, Wu et al 2018). Analysis of the record shows a variation, nominally the Hallstatt cycle, of periodicity about 2400 years. The Hallstatt cycle and a longer period cycle were also observed in $^{14}C$ record by Scafetta et al (2016). Usoskin et al (2016) concluded that the ~ 2400 year Hallstatt cycle is most likely a property of long term solar activity. They show, by superposed epoch analysis that grand minima in SSN tend to cluster, at minima of the ~2400 year cycle. The ~2400, ~ 1000, ~85, ~60 and ~ 30 year period components are important components in solar activity, cosmic ray flux and cosmogenic series and have been extensively reported (Eddy 1976, Scafetta 2010, Scafetta et al 2016, Perez-Peraza et al 2012, Usoskin et al 2016, Viaggi 2021).

In this section we show that the spectral content of $R_B/R_{SUN}$, calculated when Planet 9 is included, exhibits the Hallstatt ~2400 year cycle, the Gleissberg ~88 year cycle, the ~60 year cycle, and the ~ 30 year cycle whereas in the spectrum of $R_B/R_{SUN}$, calculated without Planet 9, these long term components of reconstructed solar activity are insignificant.

Figure 9A and 9B compare the spectral content of $R_B/R_{SUN}$ for the eight planet and nine planet systems when assessed between year -8000 and year +8000. Three overlapping nine planet spectra are shown in Figure 9A. Each spectrum is for a different value of semi-major axis and the corresponding Planet 9 orbital period. Clearly, the low frequency components of $R_B/R_{SUN}$ are sensitive to changes in Planet 9 semi-major axis and period. Therefore the variation of Planet 9 period provides a means to fit the lowest frequency component of $R_B/R_{SUN}$ to the previously observed ~2400 year period of the Hallstaat cycle in reconstructed SSN by varying the orbital radius or orbital period of Planet 9. Figure 9A indicates that a close fit is obtained with $a_9$ ~ 366 AU, or equivalently, since T and a are related, $T_9$ ~ 7000 years. The value $a_9$ ~ 366 AU is close to the best estimate value, $a_9$ = 380 AU obtained by Brown and Batygin (2021).

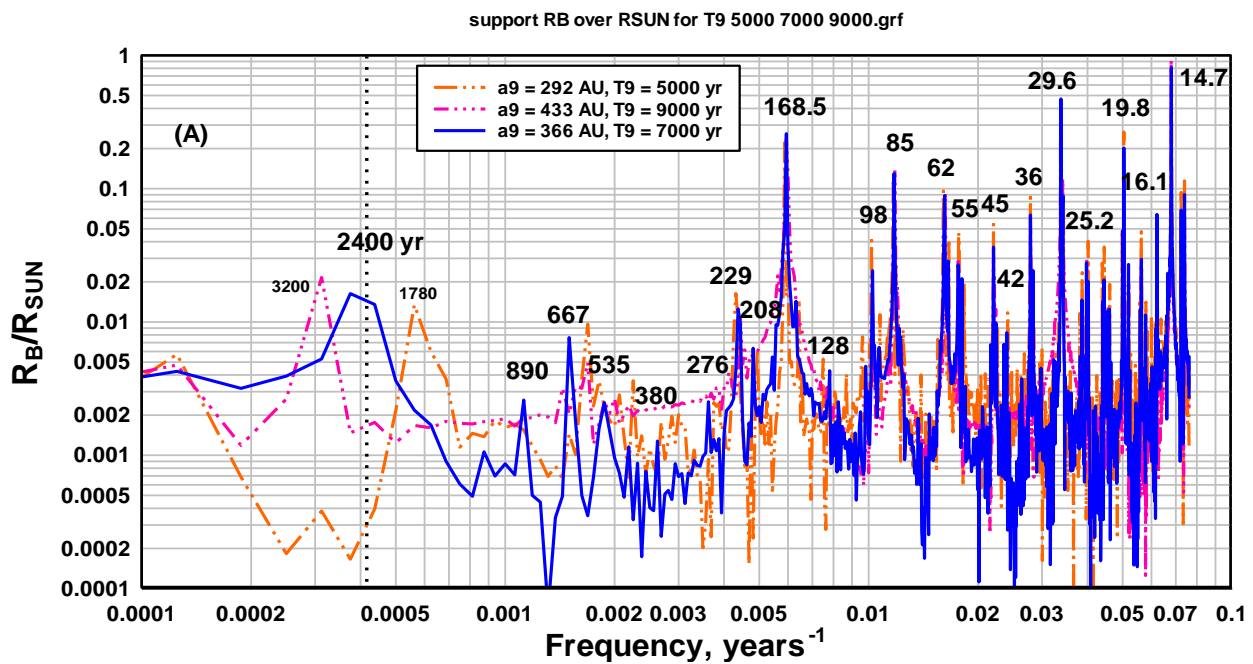

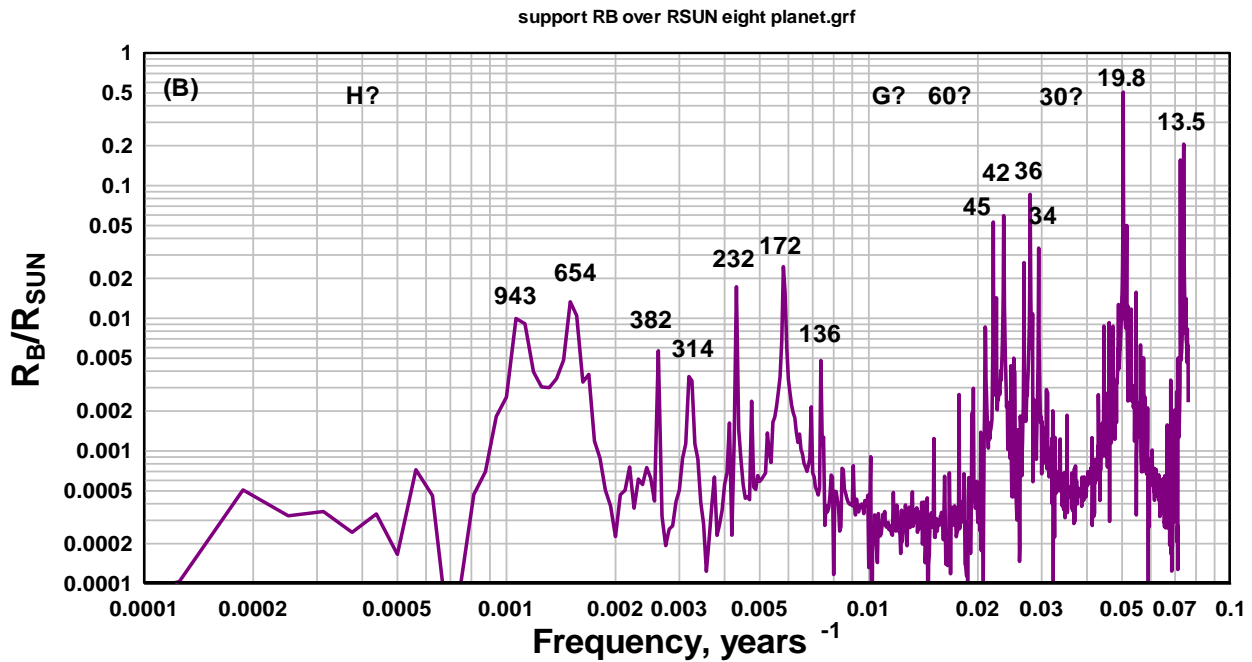

Figure 9. (A). Shows the tuning of the low frequency, long period, cycle in $R_B/R_{SUN}$ to the nominal Hallstatt period, ~2400 years, by varying the orbital period of Planet 9. (B). The components of the spectral content of $R_B/R_{SUN}$ for the eight planet system show no evidence of either the Hallstatt, ~2400 year, Gleissberg, ~ 88 year, ~60 year, or ~ 30 year period components.

Figure 10 compares the eight planet and the nine planet time variations of $R_B/R_{SUN}$ from year -8000, (8000 BC) to year +8000, (8000 AD). The points are calculated at 2400 day intervals so that the label S5 indicates a running average over a 12,000 day, 33 year interval and the label S100 indicates a running average over a 660 year interval. There is a very strong centennial time range cycle of ~ 170 year period and a weaker millennial time range cycle of ~ 2400 year period. Comparing Figure 10A where L = 60° and Figure 10B where L = 240° indicates that the ~ 2400 year cycle is sensitive to change in Planet 9 longitude, i.e. the phase of the ~2400 year cycle shifts by 180° when the longitude of Planet 9 shifts by 180°.

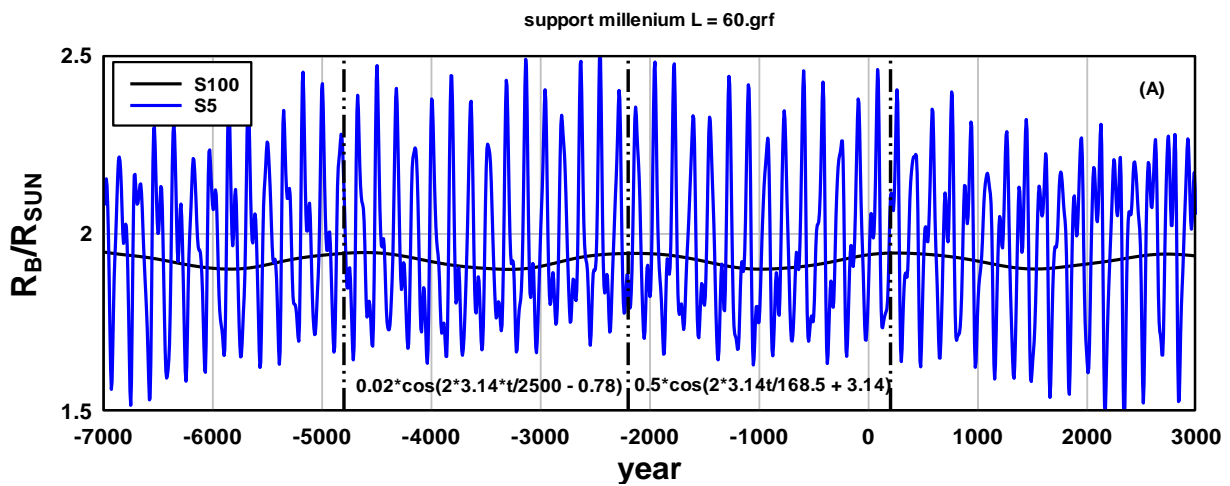

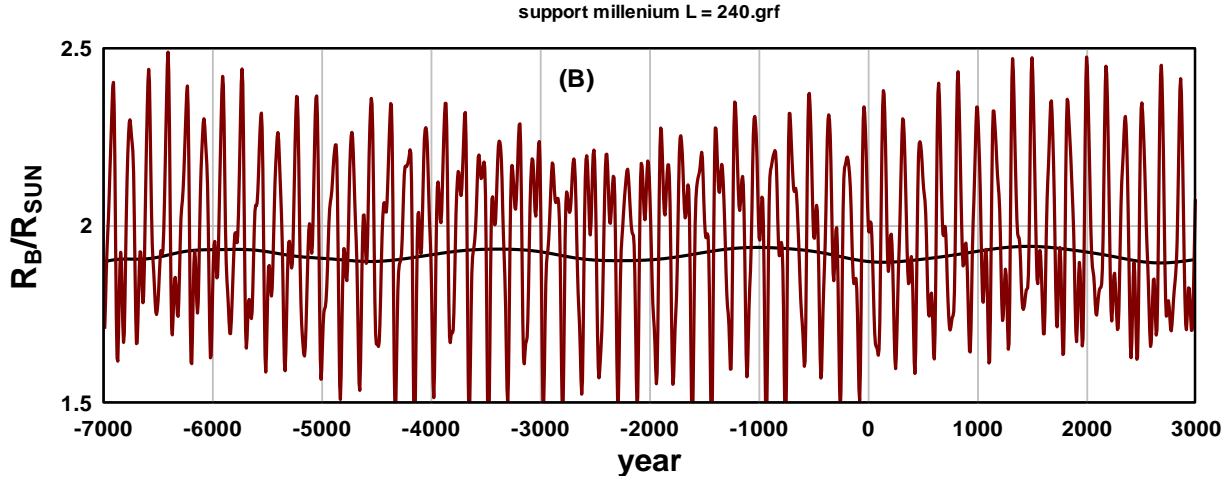

Figure 10. (A) The millennial scale variation of $R_B/R_{SUN}$ for $a_9$ = 366 AU and $L_9 = 60°$. The label S5 corresponds to a running average over 33 years and the label S100 corresponds to a running average over 660 years. The vertical reference lines in (A) correspond to times of maxima in the second singular spectral analysis component, nominally the Hallstaat cycle component, in reconstructed sunspot number, Usoskin et al ( 2016). (B) The same variations as in A but with $L = 240°$. Note that the phase angle of both the short term ~ 170 year cycle and the long term ~ 2400 year cycle shift by $180°$ when the longitude of Planet 9 changes by $180°$. The best fit to the higher frequency component in (A), nominally the Jose component, is $R_B/R_{SUN} = 0.5\cos(2\pi t/168.5 + \pi)$ and the best fit to the low frequency component, nominally the Hallstatt component is $R_B/R_{SUN} = 0.02\cos(2\pi t/2500 - 0.78)$. This component has its most recent minimum value at year 1560 in (A).

The type of variation in Figure 10 arises when the major components, in this case ~ 176 year and ~ 88 year period components are not exactly harmonically related. If the two components are harmonically related as suggested by the ratio 176/88 = 2.0 then a stationary pattern would result rather than the slowly varying pattern in Figure 10. Assuming the longer period component derives from the Uranus – Neptune conjunction period of ~ 172 years we expect the synoptic period with Planet 9 (period ~7000 years) to be 176.5 years and for the harmonic of 172 years at 86 years the synoptic period with Planet 9 to be 87.0 years. Thus Planet 9 results in a period ratio of 176.5/87.0 = 2.03 for the two components rather than 2.00 and the difference in period from harmonic ratio is sufficient to give the slowly varying pattern in Figure 10.

**5.2 The millennial variation in spectral content of reconstructed SSN.** The spectral content of the reconstructed SSN, -6755 to 1885, shown in Figure 11 was obtained after removing a running average over 2400 years from the reconstructed SSN, Wu et al (2018). The spectrum is similar to that obtained with $^{14}C$ data, Damon and Peristykh (2000).

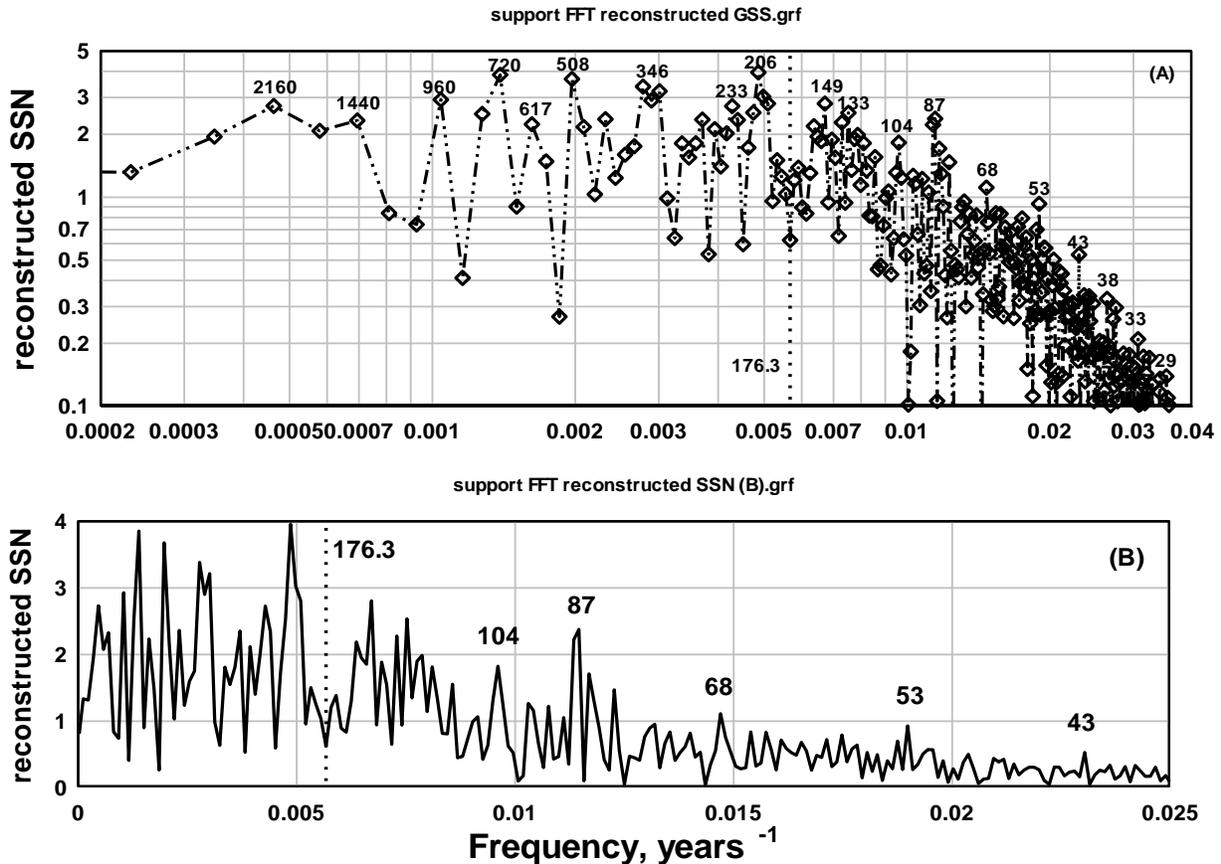

**Figure 11.** The spectral content of the reconstructed SSN record, Wu et al (2018). The period, in years, of prominent peaks are labelled. The ~ 176 year Jose periodicity, which is absent from the spectrum, is marked by the vertical dotted reference line.

It is clear by comparing the spectrum of SIM, Figure 9A and the spectrum of SSN, Figure 11, that the hypothetical transformation of SIM into sunspot emergence, if it exists, acts as a low pass filter, i.e. the transform of cycles of SIM into cycles of sunspot emergence is much stronger at low frequencies than at high frequencies. The periods of reconstructed SSN in Figure 11 along with apparently corresponding periods of $R_B/R_{SUN}$ in brackets are: 2160(2400), 1440(?), 960(890), 720(667), 617(?), 508(535), 346(380), 206(208), ???(168), 149(?), 133(?), 104(98), 87(85), 68(62), 53(55), 43(45), 38(36), 33(?), 29(29.6). Where there is no apparent correspondence a question mark is placed. Note that three question marks indicate the absence of the ~168 year Jose component from the SSN record. Clearly there is a reasonably close correspondence between the low frequency periodicity of SIM and reconstructed SSN. The notable exception is the absence of the Jose periodicity at ~ 170 years, marked with the dotted reference lines in Figure 11, from the spectrum of reconstructed SSN. This issue, which challenges the hypothesis of a connection between SIM and solar activity, is discussed in detail below.

**6. Phase modulation of the ~ 170 year Jose periodicity by lower frequency components.**

**6.1 Elements of phase modulation and demodulation.** The modulation of a high frequency signal by a lower frequency signal is the basis of communication engineering where the higher frequency signal is called the carrier and the lower frequency signal called the modulation. The basic types of modulation used in communications are amplitude modulation and phase/frequency modulation. The same concept has been applied to investigating the spectral content of solar activity and climate variables e.g. Peristykh and Damon (2003) used amplitude modulation, and (Takalo and Mursula 2002, Rial 2004) used phase/frequency modulation. In amplitude modulation a fraction of the power at the carrier frequency is shifted into sidebands on either side of the carrier frequency while in phase/frequency modulation most of the power is shifted from the carrier frequency into sidebands of the carrier frequency. This is evident, for example, in the modulation of the annual cycle of the interplanetary magnetic field by the ~ 22 year Hale cycle, where all the power in the annual cycle is shifted into sidebands, Takalo and Mursula (2002). So a characteristic of the spectra resulting from phase/frequency modulation is a distinct minimum at the carrier frequency and strong sidebands spaced equally at each side of the carrier frequency. There is a deep minimum at ~170 years in spectrum of reconstructed sunspot number, Figure 11. A possible explanation for the absence of the Jose periodicity at ~ 170 years from the spectrum is that the Jose periodicity is split into sidebands by phase modulation by lower frequency components. We investigate this possibility in the following.

Modulation of a carrier signal of angular frequency, $\omega_C$, by a lower angular frequency, $\omega_M$, modulating signal can be represented by

$y(t) = [A + \cos(\omega_M t)]\cos(\omega_C t)$

$= A\cos(\omega_C t) + \cos(\omega_M t)\cos(\omega_C t)$

$= A\cos(\omega_C t) + 0.5[\cos(\omega_C + \omega_M)t) + \cos((\omega_C - \omega_M)t)]$

$= A\cos(\omega_C t) + 0.5[\cos(\omega_H t) + \cos(\omega_L t)]$ (5)

where $\omega_H$ and $\omega_L$ are the angular frequencies of the high and low frequency sidebands respectively. When A > 1 we have amplitude modulation, A < 1 partial amplitude and phase modulation, and A = 0, pure phase modulation where on every second half cycle of the modulating signal the phase of the carrier signal is reversed. If $f_H$ and $f_L$ can be found the modulating frequency,

$f_M = (f_H - f_L)/2$ (6)

and the carrier frequency

$f_C = (f_H + f_L)/2.$ (7)

When A = 0 we have pure phase modulation. As an example, if there are two components in the modulating signal, at periods 1500 years and 600 years, and the carrier period is 168 years, the Jose periodicity, the modulated wave form and the spectrum are as shown in Figure 12 A and 12 B.

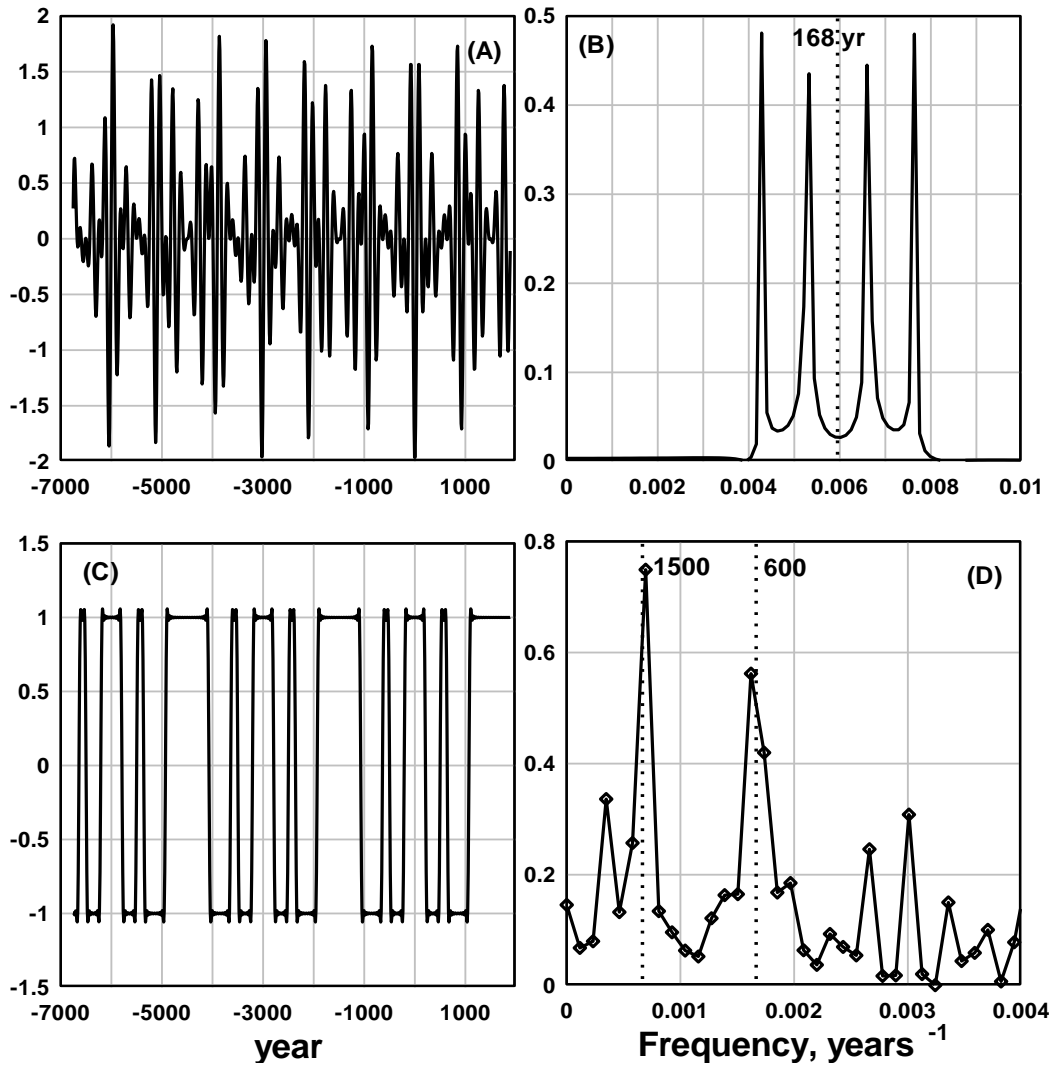

Figure 12. The figures illustrate the method of demodulating a phase modulated signal in order to reveal the low frequency modulating signal. (A) is the phase modulated signal. (B) is the spectrum of (A). The modulating periods could be obtained by applying equation 6 to find the modulating periods. However, it is possible to demodulate the signal (A) by correlating signal (A) with the carrier signal, in this simulation, $\cos(2\pi t/168)$. The result is the low frequency waveform (C), corresponding to in-phase and out-of-phase intervals of signal (A) with the carrier signal. Obtaining the spectrum of (C) by FFT recovers the periods of the low frequency modulating components, (D).

In this case, the carrier period is easy to determine as the midpoint between the sidebands in the spectrum. If the carrier period and the carrier phase is known, the low frequency modulating waveform can be recovered by demodulating the modulated signal, the variation in

(A) in the Figure 12.  Demodulation is obtained by correlating the modulated signal with the carrier signal, in the above simulation the carrier signal was cos(2πt/168 + 0). The result of the demodulation is shown in Figure 12 C which is indicative of the modulated signal alternating between in-phase and out-of-phase with the Jose cycle.  The spectrum of the demodulated signal is shown in Figure 12 D. Note that periods of the original modulating cycles, 1500 year and 600 year, are, by this process, recovered from the modulated waveform. Clearly the demodulation method is not exact and weak spurious components are generated as evident in Figure 12 D. It is also possible, by using slight variations in period and phase of the Jose cycle, and the above simulation methods, to show that the demodulation method is very sensitive to the period and phase of the Jose cycle.

**6.2  Phase modulation and demodulation applied to the reconstructed sunspot number**  It is clear from the above that it is possible to move back and forth between a modulated signal and the modulating signal. We now apply this method to the reconstructed SSN record with the objective of demonstrating the possibility that the Jose cycle is the "hidden" mediator between the low frequency and mid frequency components of the SSN record. Figure 13 outlines the transformation by which the low frequency components of the SSN variation modulate the Jose cycle and result in the mid frequency components of the SSN.

The very low frequency component of SSN record, period longer than ~2400 years, that Usoskin et al (2016) regarded as likely being of terrestrial origin, was removed by subtracting a 2400 year running average. This resulted in the very complex blue broken curve shown in Figure 13 A.  A running average over 30 points, or 300 years, was then used to obtain the low frequency waveform, the full line in Figure 13 A. The spectrum of this low frequency SSN waveform, shown in Figure 13 B, has seven significant low frequency components. Figure 13 C is the result of modulating the Jose cycle, cos(2πt/168 + π), obtained from Figure 10A, with the low frequency waveform of Figure 13 A.  The resulting spectrum, corresponding to variation in the mid frequency range of the SSN record is shown in Figure 13 D, the most noticeable feature being the distinct minimum in spectral amplitude centred on the Jose period, 168 years. This spectrum, Figure 13 D, is directly comparable, in the same frequency range with Figure 11 B, which shows the entire SSN spectrum.  The sidebands in Figure 13 D are the result of phase modulation of the Jose cycle by the low frequency components in Figure 13 A. For example the most central pair of sidebands is due to modulation by the 0.0005 year$^{-1}$ , ~ 2000 year, component in Figure 13 A and the second set of sidebands from the centre is due to phase modulation by the ~ 1000 year component in Figure 13 A.

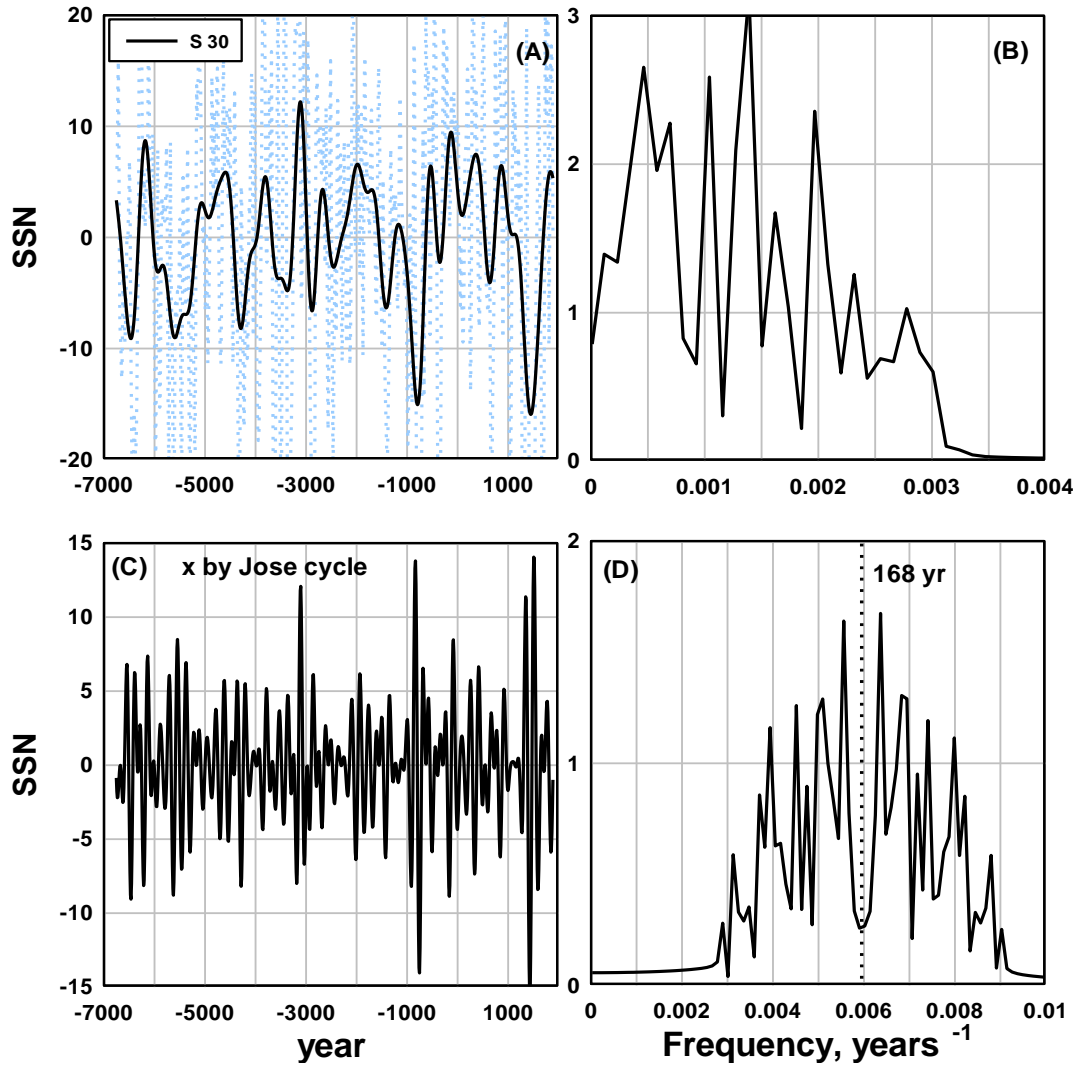

Figure 13. The figures illustrate the modulation of the Jose cycle by the low frequency components in the reconstructed SSN record. The blue broken curve in (A) is the time variation of the reconstructed SSN record -6755 to 1885, Wu et al (2018). The full line is the low frequency component obtained by a 300 year running average. The spectrum of the low frequency SSN variation is shown in (B). If the low frequency time variation in (A) modulates a Jose cycle, $\cos(2\pi t/168 + \pi)$ the resultant mid frequency SSN variation is shown in (C). The spectrum of this mid frequency variation shown in (D) is characterised by a deep minimum centred on the Jose periodicity, comparable with the spectrum of Figure 11 B. The sidebands correspond to phase modulation of the Jose cycle by the low frequency components in (A).

We now demonstrate the demodulation of the SSN record in the mid frequency range via correlation with the Jose cycle to obtain the low frequency components of the SSN record. The mid frequency range variation of SSN record is obtained by using a digital band pass filter to isolate the variation in the frequency range from 0.003 years$^{-1}$ to 0.009 years$^{-1}$, Figure 14 A. The spectrum of this variation, shown in Figure 14 B, has all the low frequency components of the SSN variation removed by the filter. We now correlate the mid frequency range variation,

Figure 14 A, with the Jose cycle, $\cos(2\pi t/168 + \pi)$, as determined in Figure 10A. The variation of the resulting correlation coefficient, after smoothing with a 100 year running average, is shown in Figure 14 C. The spectrum of this variation, shown in Figure 14 D, contains low frequency components similar in period to the low frequency components of the SSN record in Figure 13 B.

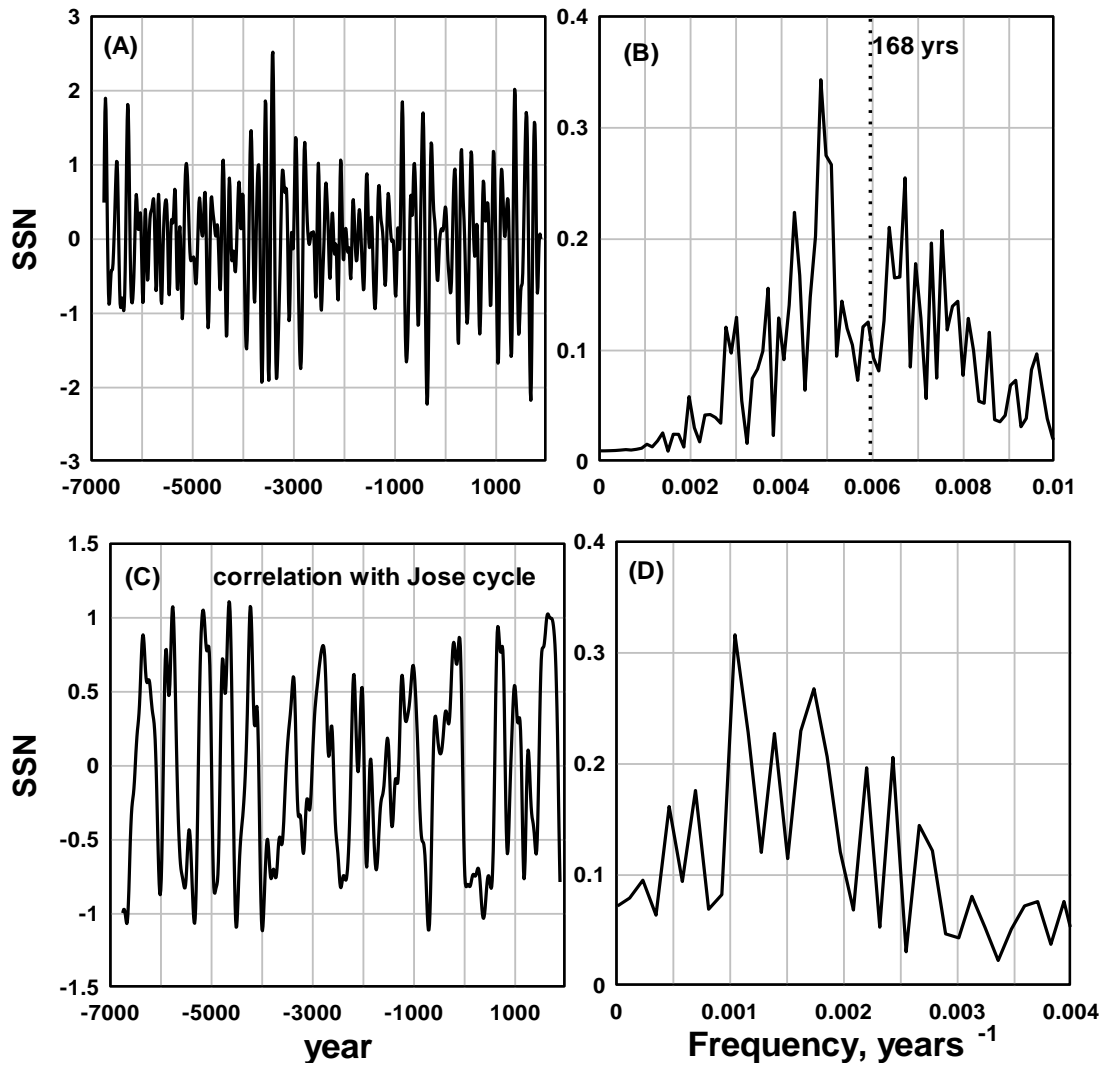

Figure 14. The figures illustrate the demodulation of the mid-frequency SSN record to obtain the low frequency components of the SSN record. The variation in (A) is the mid-frequency variation of the SSN record obtained by digital band pass filtering. The spectrum (B) is the result of a FFT of variation (A) and shows the deep minimum at about 0.006 years$^{-1}$ characteristic of phase modulation of a component at that frequency. (C) shows the correlation of the variation in (A) with the Jose cycle $\cos(2\pi t/168 + \pi)$ after smoothing with a 100 year running average. (D) is the FFT of (C) and corresponds to the low frequency components of the SSN record and is directly comparable with **Figure 13 B**.

The variation of Figure 14 C and the spectrum of Figure 14 D are reasonably consistent with the low frequency variation of Figure 13 A and the low frequency spectrum of Figure 13 B, indicating that the demodulation of the mid frequency variation in the SSN record with the Jose cycle is recovering, to a reasonable approximation, the low frequency variation in the SSN record. The effectiveness of the type of transformations illustrated in Figures 13 and 14 support the concept that phase modulation of the Jose cycle in SIM by the low frequency components in SIM results in the mid frequency variation of the reconstructed SSN. It also explains the occurrence of the deep minimum in the spectrum of the SSN record at ~ 170 years period that is evident in the spectrum of reconstructed SSN, Figures 11 A and 11 B, and in previous observations of the spectral content of cosmogenic isotope series, (Peristykh and Damon 2003, Knudsen et al 2009, Viaggi 2021). Note that in a later section, Section 6.4, we find that the Jose periodicity in SIM varies slightly with Planet 9 orbital radius and, if the Planet 9 orbit is eccentric, an exactly constant Jose periodicity may not apply to SIM.

The hypothetical connection between SIM and SSN is that SIM, by some mechanism, as discussed below, induces changes in the magnetically active region of the Sun that vary the rate of emergence of sunspots. The connection between SIM and the reconstructed SSN involves several further connections. Connections between sunspots and solar wind, solar wind and cosmic rays, cosmic rays and the production of $^{14}$C or $^{10}$Be isotopes, the accumulation of the isotopes in tree rings or ice layers, the interpretation of tree rings and ice layers into a time sequences, and the conversion of the resulting cosmogenic time series into a proxy record of sunspot emergence, Wu et al (2018). In view of the noise, non-linearities and uncertainties involved in such a complex process it is unlikely that the analysis, outlined above, based on phase modulation of the Jose cycle, Figures 13 and 14, would yield a very close resemblance between the observed and estimated low and mid frequency components of the reconstructed SSN. Nevertheless the comparison is, on close examination, reasonably convincing and we conclude, tentatively, that in the transformation from SIM to solar SSN, the Jose cycle in SIM is phase modulated in the process and that this phase modulation could be the reason for the absence of Jose periodicity in the spectra of SSN.

**6.3 Spectral analysis of SSN based on the dates of Grand Solar Minima** Grand solar minima have been a preoccupation of solar and climate researchers since the reporting of the Maunder solar minimum by Eddy (1976). The study of grand solar minima is important in the context of understanding the origin of long term variations in the solar dynamo, (Charbonneau 2020, Inceoglu et al 2016, Inceoglu et al 2017, Usoskin 2016, Cionco and Soon 2015), and in relation to the possible climate impacts of grand solar minima, e.g. (Lockwood et al 2010, Feulner and Rahmsdorf 2010, Meehl et al 2013, Ineson et al 2015, Viaggi 2021). The cosmogenic record has enabled the study of the occurrence of grand solar minima over the last ten millennia with Usoskin et al (2007) reporting the centre times of grand minima in the reconstructed SSN and Inceoglu et al (2015) reporting the centre times of grand minima in the solar modulation potential. The conclusions of the latter two studies is that, "the occurrence of grand minima and maxima is not a result of long-term cyclic variability but is defined by stochastic/chaotic processes" and "other than a weak quasi-periodicity of 2000-2400 years ….. No other periodicities are observed in the occurrence rate of grand minima", Usoskin et al (2007). The conclusions are counter to the findings in this section that grand solar minima result from the ~ 170 year Jose cycle phase modulated by lower frequency cycles such as the ~ 1000 year Eddy and ~ 2400 year Hallstatt cycles. However, Inceoglu et al (2016) have changed their view about grand solar minima to, "it would also imply that these quiescent periods of activity are not the result of a random process but instead their origin is linked to the driving mechanism of magnetic field generation".

With a view to assessing if long term cyclic variability is present or not in the occurrence of grand minima we here study the central dates of grand minima as provided by Usoskin et al (2007) and Inceoglu et al (2015). The objective is to assess if the ~ 170 year Jose cycle and the harmonic and sub harmonics of the Jose cycle are evident in the occurrence dates.

The method is to specify, in the relevant time interval, a value unity for the centre years of grand solar minima and a value zero for all other years. To enable the use of FFT analysis the central dates of grand solar minima are assigned to the nearest fifth year in the sequence of years between -9200 and 1810 and given the value unity. The sequence, after smoothing with a ten year running average, and representing the occurrence of grand minima as specified by Usoskin et al (2007) or Inceoglu et al (2015), is shown in Figure 15A for the Usoskin et al (2007) data and Figure 16A for the Inceoglu et al (2015) data. A fast Fourier transform of the sequence is then made to identify periodicities in the sequences, Figure 15B and Figure 16B. This method is similar to that applied to uncover the spectral content of ground level enhancement events, Velasco Herrera et al (2018).

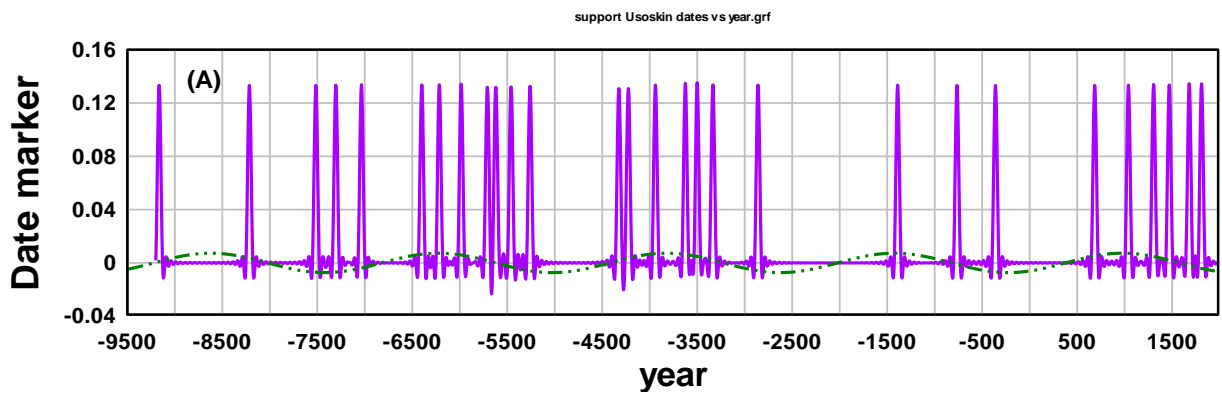

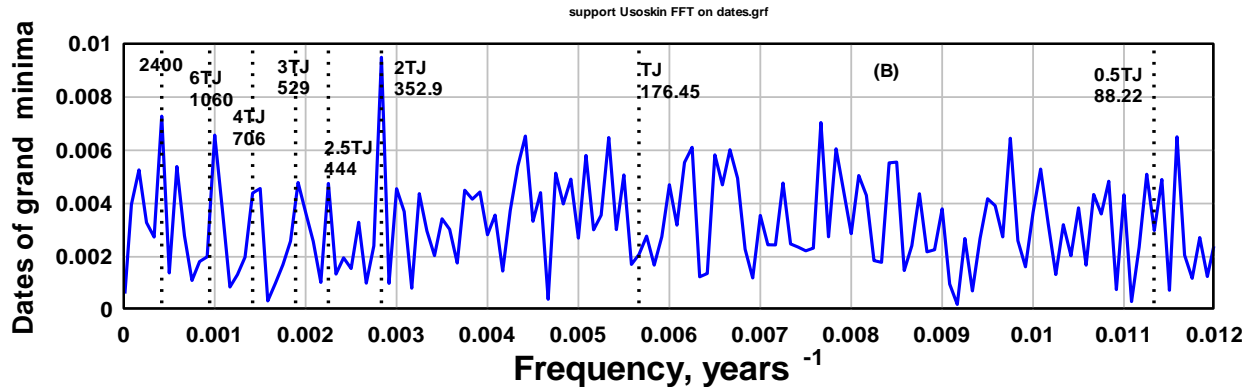

Figure 15. (A). The time sequence of the central dates of solar grand minima, Usoskin (2007). (B) The spectral content of the time sequence in (A). The strongest peak, at 352.9 years, is assigned the value 2TJ, where TJ is the Jose period in SSN. TJ is then calculated to be 176.45 years and this period is indicated by the TJ reference line. The harmonic and sub harmonics of the Jose components are then indicated by reference lines. A peak at 2400 years is also indicated.

The strongest peak in the spectrum of the Usoskin et al (2007) data is at 352.9 years. We associate that period with 2TJ where TJ is the Jose periodicity, 176.45 years. As expected from the discussion about phase modulation, a prominent peak at TJ is not evident in the spectrum and the period TJ is associated with a deep broad minimum in the spectrum. Four sub-harmonics, 2.5TJ, 3TJ, 4TJ, and 6TJ are at prominent peaks. The first harmonic of the Jose periodicity at 0.5TJ = 88.22 years is also indicated. The long term periodicity at ~2400 years that represents the clustering of solar grand minima occurrences is also marked. Note that the ~ 2400 year cycle of clusters in Figure 15A corresponds to the negative phase of the ~ 2400 year Hallstatt cycle in Figure 10A.

The analysis is repeated for the Inceoglu et al (2015) reconstructions of solar modulation potential from the $^{10}$Be and $^{14}$C cosmogenic series. The sequence of grand solar minima occurrences is shown in Figure 16A and the resulting spectral content of the occurrences shown in Figure 16B. The periodicities marked are essentially the same as in the spectrum of Figure 15B derived from the Usoskin et al (2007) data. Thus three different estimates of the occurrence of grand minima give essentially the same periodicity of occurrence and that periodicity is dominated by a Jose periodicity of ~ 177 year period.

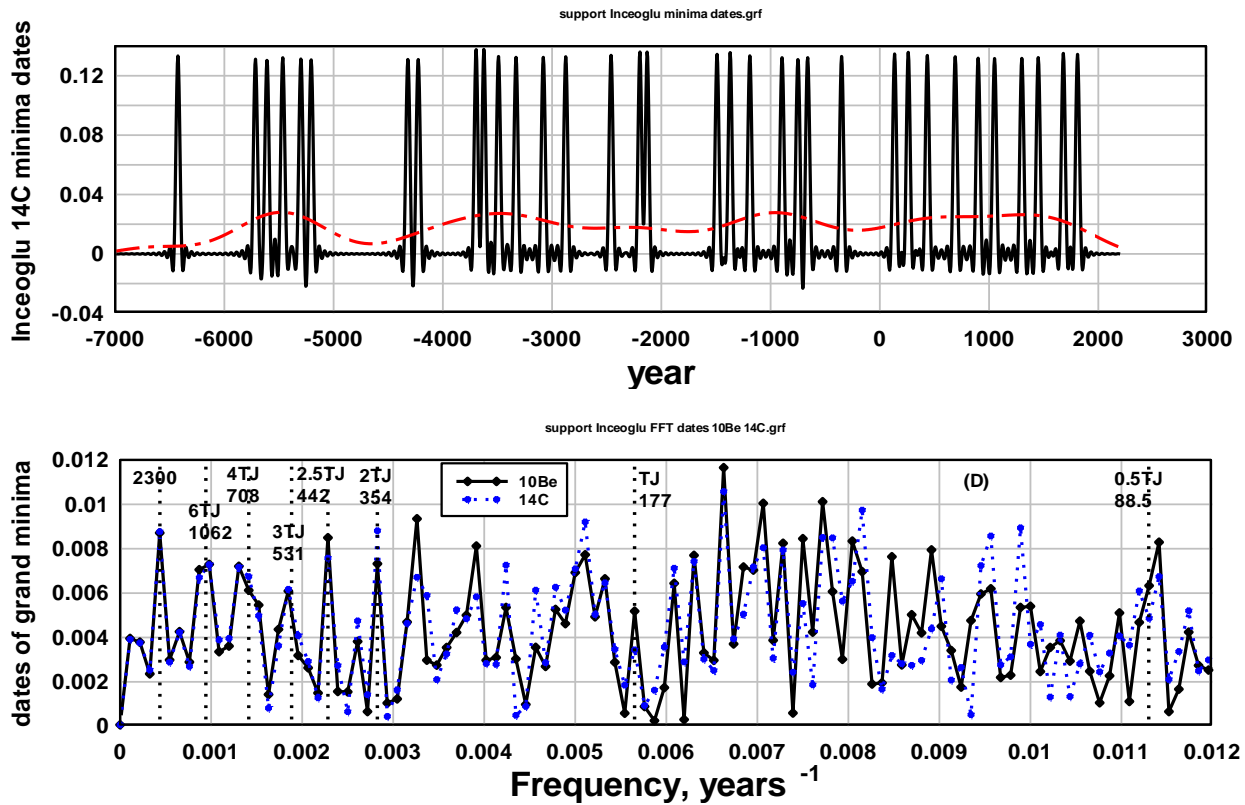

**Figure 16.** Upper graph. The time sequence of the central dates of solar grand minima, Inceoglu et al(2015) in the $^{14}$C record of solar modulation potential. Lower graph. The spectral content of the time sequences in 14C and 10Be solar modulation potential. The strong peak, at 354 years, is assigned the value 2TJ, where TJ is the Jose period in SSN. TJ is then calculated to be 177 years and this period is indicated by the TJ reference line. The harmonic and sub harmonics of the Jose components are then indicated by reference lines. A peak at 2300 years is also indicated.

The ~177 Jose periodicity in the occurrence of solar grand minima, while clearly associated with the low frequency components in the spectrum is "hidden" in the sense that the amplitude of the component is relatively weak compared with the amplitude of the sidebands and with the amplitude of low frequency components. This is a result, as outlined in section 6.2, of phase modulation of the Jose component by lower frequency components. However, the phase modulated component is not completely "hidden" if the number of phase reversed cycles or events differs significantly from the number of regular cycles or events. In the context of grand solar minima occurrences, if all grand minima occurred at n x 176 year intervals in the time sequence then there would be a strong peak at 176 years in the spectrum of occurrences. If, however, about half of the grand minima occurred at (n + ½) x 176 year intervals in the time sequence the 176 year periodicity in the spectrum would be "hidden". In the spectrum of Figure 16 B a peak at ~ 177 years is evident but relatively weak indicative of an imbalance between in phase and out-of-phase occurrences of grand solar minima. There are several strong peaks in the spectrum that correspond to phase modulation of the Jose cycle that have not been marked by reference lines. For example, using equations 6 and 7, the peaks at 303 years, (0.0033 years$^{-1}$), and 256 years, (0.0039 years$^{-1}$), in Figure 16 B are sideband peaks due to phase modulation of the Jose cycle by the 442 year period and 531 year period low frequency components respectively. Similarly the strong, broad peaks at ~ 0.005 year$^{-1}$ (200 years) and ~0.007 years$^{-1}$ ( 143 years) are sideband peaks due to phase modulation of the ~ 177 year Jose cycle by the ~ 1000 year Eddy cycle.

Beer et al (2018) suggested that solar grand minima "recur with the characteristic de Vries period of approximately 208 years" and supported this idea by band pass filtering the cosmogenic $^{14}$C record in the period range between 180 and 230 years, further stating that "the maxima in the filtered cosmogenic record "actually correspond to grand minima in solar activity". The evidence of Figure 16B suggests that the recurrence time of solar grand minima obtained from the $^{14}$C record is close to 177 years. If the recurrence time was 208 years a peak at 0.0048 years$^{-1}$ would be evident in Figures 15B and 16B.

The above analysis is presented as a possible reason why the Jose periodicity, a dominant periodicity in the spectrum of SIM, is absent from the spectrum of reconstructed SSN whereas other periodicities in SIM, e.g. the ~ 2400 year Hallstatt, ~ 1000 year Eddy, and the ~ 88 year Gleissberg periodicities, are clearly present in the spectrum of reconstructed SSN, c.f. Figures 11A and 11B.

Another conundrum in this attempt to relate SIM to SSN is the discrepancy between the ~ 177 year Jose periodicity obtained by spectral analysis of the solar activity record, Figure 11A and Figure 16 B, and the ~ 170 year Jose periodicity obtained from SIM, Figure 9A and Figure 10A. Sharp (2013) and McCracken et al (2014), using very long records of SIM estimated the Jose period to be ~ 171 years as opposed to the estimate by Jose (1965), over the short interval 1653 to 2060, of ~ 178 years. The estimate obtained in this paper using the approximation of circular planet orbits is ~ 169 years for SIM including Planet 9 and ~ 172 years for SIM for the eight planet system, the latter value consistent with the estimate by Sharp (2013) and McCracken et al (2014). The discrepancy between ~ 177 years for SSN and ~ 170 years for SIM is large enough to be a significant challenge to the planetary hypothesis.

**6.4 Why does the Jose periodicity in SSN differ from the Jose periodicity in SIM?** The previous section found the "hidden" Jose periodicity in SSN is ~ 176.5 years, Figures 15 and 16, whereas the calculated Jose periodicity in SIM was ~ 168.5 years, Figure 9A. This discrepancy may be resolved by noting that the Jose periodicity in SIM varies with the semi-major axis, $a_9$, of Planet 9, Figure 17. The range is fairly wide. When $a_9$ = 150 AU the Jose period is 182 years and when $a_9$ = 800 AU the Jose period is 165.9 years. In fact all of the SIM periodicities that depend on the semi-major axis of Planet 9 vary. The Geissberg ~ 88 year periodicity in SIM also decreases with $a_9$, Figure 17. However, in contrast, the Hallstaat periodicity in SIM increases strongly with $a_9$, Figure 9A.

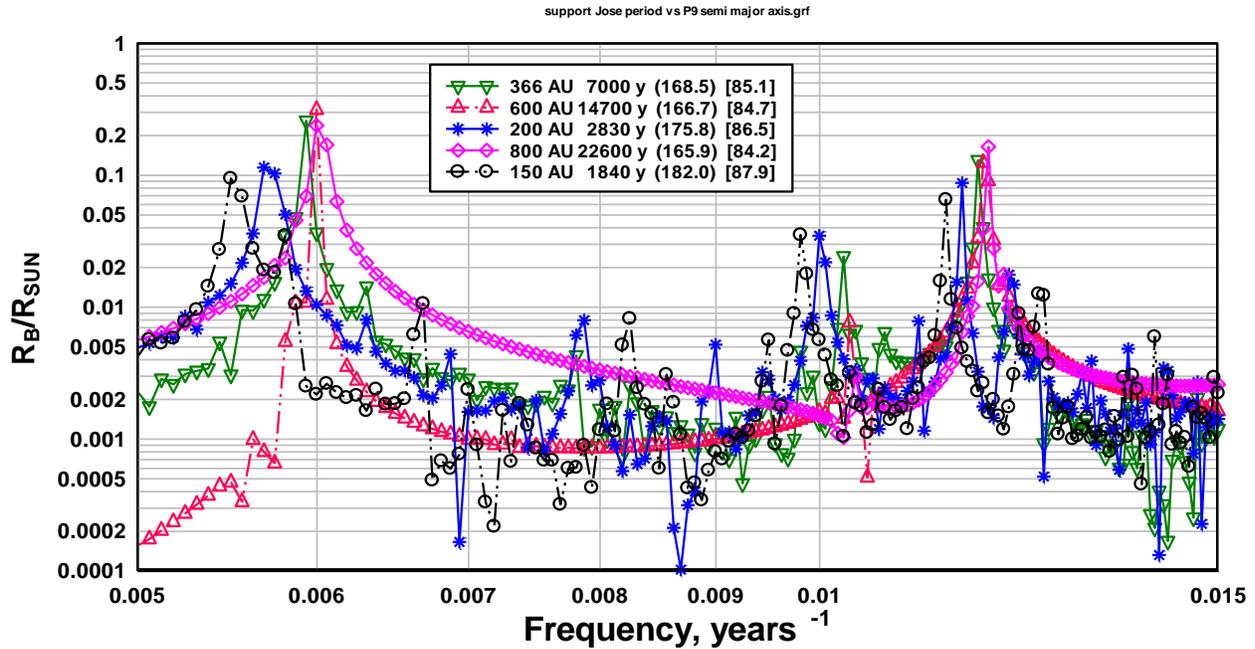

Figure 17. The spectral content of RB/RSUN as the semi-major axis of Planet 9 is varied between 150 AU and 800 AU. Note that all the periodicities move to lower frequencies as the Planet 9 semi-major axis decreases. The semi-major axis and orbital period of Planet 9 are given in the lgend as is the period, in years, of the Jose cycle in SIM, indicated (….), and the period of the Gleissberg cycle in SIM, indicated [….].

Thus, a possible resolution of the difference between the SSN and SIM estimates of Jose periodicity is that the orbital radius of Planet 9 used in the calculation of SIM in this paper, 380 AU, the best estimate by Brown and Batygin (2021), is actually shorter, about 250 AU. Another possibility is that the Planet 9 orbit is eccentric of the order e ~ 0.4, and the occurrence of grand solar minima cluster around the times of perihelion when the orbital radius approaches a lower value of 250 AU. In this case most of the solar grand minima in SSN would be expected to occur, in clusters, at a time spacing of about 176 years, as observed, corresponding to the Jose periodicity in SIM when $a_9$ ~ 250 AU, also ~ 176 years.

## 7. Mechanism for the conversion of SIM into SSN.

There is a conundrum that, while there is reasonably close correspondence between most of the significant periodicities in SIM and in SSN, the principal component in SIM, the ~ 170 year Jose periodicity, is not evident in spectra of reconstructed SSN or in cosmogenic records, Viaggi (2021). The inclusion of Planet 9 in the solar system results in broad decreases in $R_B/R_{SUN}$ coherent with the occurrence of grand solar minima, c.f. Figure 1 and Figure 7. A significant long term variation in Sun to barycentre distance implies that the Sun is subject to intervals of extreme positive and negative acceleration with the intervals separated by ~ 170 years. Figure 18 is the second time differential of the $R_B/R_{SUN}$ variation in Figure 7. The broken vertical reference lines are the times centred on the six solar grand minima referenced in Figure 7. The noticeable feature is that times of the centres of grand minima coincide closely with extremes of negative to positive solar inertial acceleration. The acceleration indicated by the level 0.001 in Figure 18 is equivalent to $0.7 \times 10^{-9}$ m/s$^2$. This is a small acceleration, about the same as the tidal acceleration on the Sun due to Jupiter, de Jager and Versteegh (2005), however, it acts over a long time. For example, during the positive acceleration peaks during the Sporer, Maunder and Dalton grand solar minima the acceleration exceeds this level for ~ 20 years.

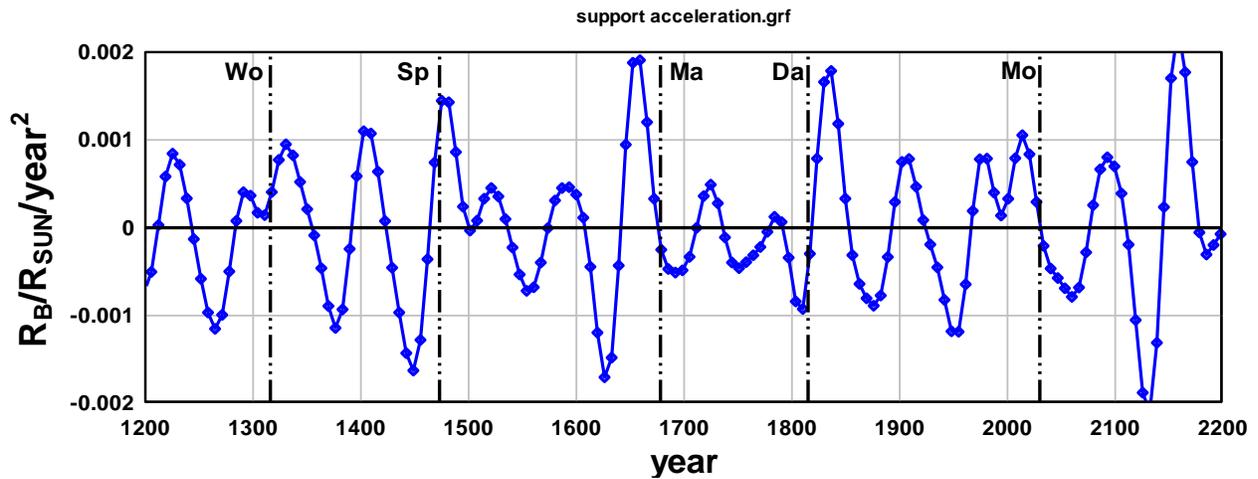

Figure 18. The acceleration along the Sun – barycentre direction due to planet motion including Planet 9. This curve is the second differential of the variation with $L_9 = 60°$ in Figure 7. The reference lines, aside from the line for the Modern grand minimum, mark the centre of grand solar minima based on the data of Usoskin et al (2021). The reference time for the Modern Minimum is an estimate. The 0.002 level in the figure is equivalent to a Sun acceleration of $1.4 \times 10^{-9}$ m/s$^2$.

The inclusion of Planet 9 provides for significant acceleration of the Sun back and forth along the line between the Sun and barycentre in the time range of grand solar minima. This suggests a mechanism for SIM influencing the decrease in solar dynamo action during grand solar minima.

The Sun has a spherical solid core extending to a radius of $0.7R_{SUN}$ separated from a spherical fluid convective outer layer $0.3R_{SUN}$ thick containing about 2% of the Sun's mass. This is the region where meridional and differentially rotating flows of material occur. When the Sun is accelerating along the line connecting the Sun to the barycentre the fluid convective region is displaced relative to the inner solid core and the convective region bulges in the acceleration direction and thins in the two orthogonal directions, in particular in the directions perpendicular to the ecliptic plane, the polar directions. Therefore the result of positive or negative solar acceleration is a thinning of the convective region at both poles.     Simple hydraulic considerations indicate that the meridional and differential rotation flows would be slower in regions where the convective layer is thinner and as a result magnetic activity and sunspot emergence would decrease. In support of this idea, the impedance to hydraulic flow between two parallel planes varies approximately as $1/d^3$ where d is the distance between the planes, the meridional flow converges at the poles, and the thinner convective region near the poles has to accommodate meridional flows towards and away from the poles; all suggesting strong dependence of meridional flow on the thickness of the convective layer at the poles. Simulation of solar magnetic fields during the Maunder Minimum, (Wang and Sheeley 2003, Karak 2010, Karak and Choudhuri 2013, Choudhuri 2020, Miyahara et al 2021), indicated that the Schwabe sunspot cycle was extremely sensitive to meridional flow speed. The mechanism proposed here provides simultaneous thinning at both poles so that both northern and southern hemisphere meridional flows would be reduced. This would result in symmetric decrease in SSN in both hemispheres. In contrast, the assumption that only stochastic processes affect the meridional flows leads to highly asymmetric SSN variation, Brandenburg and Spiegel (2008), with decreased SSN in one hemisphere and normal SSN in the other hemisphere. Direct measurement of meridional flow has been possible only since 1996, (Hathaway and Rightmire 2010, Mahajan et al 2021) with studies finding that a broad reduction in meridional flow speed of about 5 m/s occurred with the reduction centred on year 2000. This direct observation of meridional flow speed reduction corresponds with the broad peak of Sun-barycentre acceleration centred on year 2000 in Figure 18 and provides evidence supporting the relationship proposed for the mechanism. We note from Figure 18 that moderate SIM accelerations occur ~ 2050 and ~ 2090. However, extreme acceleration, equivalent to that associated with the maunder minimum will not occur until 2130.

The problem faced when proposing the type of mechanism outlined above, and for most mechanisms based on planetary influence on the Sun, is showing that the planetary accelerations are sufficient to cause or trigger variation in solar flows large enough to affect the solar dynamo. For example, it is known that by simulating a slow random variation on meridional flow in the convective layer, it is possible to simulate grand minima type variations of the solar dynamo, (Passos and Lopes 2008, Karak 2010, Karak and Choudhuri 2013). However, the required reduction in meridional flow was found to be about 5 m/s. Whether accelerations of the Sun the order $10^{-9}$ m/s$^2$ acting over ~20 years can produce that magnitude of change in the meridional flow is outside the scope of this paper. However, the model outlined above is a mechanism of planet to Sun influence that qualitatively explains many of the observations associated with grand solar minima. Some examples are discussed below.

One example is the occurrence of a normal solar cycle near the middle of solar grand minima. As indicated in Figure 18 the observed centre times of grand solar minima occur near the times of extreme positive or negative inertial acceleration and the time interval between the acceleration extremes is commensurate with the duration of grand solar minima, ~ 50 years, Inceoglu et al (2015). At the changeover between positive and negative acceleration the solar acceleration is close to zero and meridional flows and the solar dynamo should return to normal function during these short, approximately decade long, intervals. Thus a cycle of near normal amplitude and length would be expected to occur around the middle of grand solar minima. The analysis of reconstructed SSN over the last millennium, Usoskin et al (2021), shows a single, relatively strong, and short cycle occurs near the middle of the Wolf, Sporer and Maunder grand minima, with cycle maximum at years 1313, 1494, and 1682. It is clear from Figure 18 that these times are coincide with times when solar acceleration is close to zero.

A second example is the provision of a mechanism by which the ~ 170 year Jose cycle in SIM is phase modulated in the transformation of SIM into SSN. An explanatory mechanism is important as the ~ 170 year cycle is very prominent in SIM spectra yet is weak, absent, or as discussed above "hidden", in the spectra of SSN and cosmogenic records as reported, for example, by McCracken et al (2014) and Viaggi (2021). McCracken et al (2014) discussed this issue and concluded that the non-appearance of the ~ 170 year periodicity in SSN spectra is "a consequence of the cycle to cycle variability of the Jose cycle as a consequence of the non-commensurate nature of the periods of the Jovian planets". However, this is a general statement of the problem and not an explanation of it. The mechanism outlined above can explain how the Jose cycle in SIM results in a phase modulated variation in SSN. The basis of a phase modulation mechanism is that an event in solar activity, e.g. the occurrence of a solar grand minimum, should sometimes occur during a positive phase of the Jose cycle in SIM and at other times occur during the negative phase of the Jose cycle in SIM. The Jose cycle in SIM for the last millennium is illustrated in Figure 7. Here we see that the occurrence of grand minima in SSN occur close to sharp turning points in SIM during the negative phase of the Jose cycle. The sharp turning points in $R_B/R_{SUN}$ correspond to times of extreme solar acceleration, and, on the basis of the proposed mechanism, to a reduction in the solar dynamo. Now turning to Figure 10A and the variation of SIM in the interval between -7000 to 3000 we notice that, during the interval 0 to 3000, SIM is characterised by sharp negative excursions during the negative phase of the Jose cycle and similarly from -7000 to -5000. However, during the interval -5000 to 0 the SIM is characterised by sharp positive excursions during the positive phase of the Jose cycle. In the mechanism proposed here grand solar minima occur near the times of extreme solar acceleration due to thinning of the convective layer. It follows that during the interval – 5000 to 0 grand minima in SSN will occur, predominantly, during the positive phase of the Jose cycle in SIM, whereas, during the intervals -7000 to -5000 and 0 to 3000 grand minima in SSN will occur, predominantly, during the negative phase of the Jose cycle in SIM. As the latter two intervals encompass ~ 5000 years, and the interval -5000 to 0 encompasses ~ 5000 years, we would expect approximately equal occurrence of grand minima in SSN during the positive and during the negative phases of the Jose cycle in SIM. This would result in the appearance of phase modulation of grand minima in SSN with the centre period of the modulation at the period of the Jose cycle in SIM. Thus, the proposed mechanism resolves the conundrum of the strong Jose component in SIM being an absent or a "hidden" component in SSN as outlined in section 6 of this paper. Occasionally moderately sharp turning points in SIM and accompanying moderately strong accelerations of SIM will occur during the broader intervals in the variation in SIM. For example, the moderately sharp turning points within broad intervals that occur near year 1400 and near year 1910 in Figure 7, would also result in reductions in SSN. These moderate solar minima, occurring at half intervals of the Jose cycle, would also lead to phase modulation in SSN for similar reasons as discussed above.

While the mechanism described above is mainly qualitative it has, as it is based on planet motion, a predictive capacity. For example, Figure 7 and Figure 15 relate the ~ 170 year variation in SIM quite accurately to the times of occurrence of the grand minima of SSN in the cluster of grand minima occurring in the last millennium. Further, the occurrence of the cluster can be accurately related to the SIM estimate of the time of occurrence of the minimum of the ~ 2400 year Hallstatt cycle during the last millennium, at ~ year 1500, Figure 10. However, it is clear that the occurrence of solar grand minima is not only due to modulation of the Jose cycle by the Hallstaat cycle but also to the modulation of the Jose cycle by several other low frequency cycles, c.f. Figures 11, 13, and 14. Thus, predicting the occurrence of future grand minima would require knowledge of the phases of the other low frequency cycles as well as the phase of the Hallstaat cycle. It is useful to note that, for dynamo theory based entirely on stochastic simulation of meridional flows, (Wang and Sheeley 2003, Karak 2010, Charbonneau 2020, Karak and Choudhuri 2013, McIntosh and Learmon 2015), while it is possible to simulate grand solar minima, it is not possible to make predictions of when grand solar minima in SSN will occur. As Wang and Sheeley (2003) note, "the origin of the assumed fluctuations is unknown to us" and as Karak (2010) noted, "We have no idea why the meridional circulation dropped to a very low value. ……. However, this assumption enables us to reproduce most of the important features of the Maunder minimum remarkably well", and as Charbonneau (2020) stated " At this writing we still do not know what triggers Grand Minima, or which physical processes control their duration and drive recovery to "normal" cyclic activity".

## 8. Conclusion

The main hypothesis of this work is that including Planet 9 in the solar system improves the coherence between SIM and SSN. This hypothesis was supported by the demonstration that the important cycles in SSN, the ~ 2400 year Hallstaat, the ~ 88 year Gleissberg, the ~ 60 year, and the 30 year cycles, emerged strongly in SIM when Planet 9 was included in the calculation of SIM. The increase in spectral content with Planet 9 included, c.f. Figures 9A and 9B, is important as it has been suggested, Callebaut et al (2012), that identifying the principal periodicities observed in solar and climate variability is one of the two crucial tests of whether there is a planetary influence on the solar dynamo and the variation solar activity.

With the development of a mechanism for the effect of SIM on SSN, i.e. that solar acceleration thins the convection layer at the poles thereby reducing meridional flows and reducing sunspot activity, it was possible to explain the occurrence of grand solar minima and provide an explanation why the Jose cycle is a hidden periodicity in the spectra of SSN records.

The initial finding of the discrepancy between the Jose periodicity in SIM, ~ 170 years, and the Jose periodicity in SSN, ~ 177 years was resolved by evidence that the Jose periodicity in SIM varies with Planet 9 orbital radius over the range that includes the period 177 years.